# pyBibX - A Python Library for Bibliometric and Scientometric Analysis Powered with Artificial Intelligence Tools

*Valdecy Pereira*
*valdecy.pereira@gmail.com*
*Federal Fluminense University*

*Marcio Pereira Basilio*
*marciopbasilio@gmail.com*
*Federal Fluminense University*

*Carlos Henrique Tarjano Santos*
*tesserato@hotmail.com*
*Federal Fluminense University*

**Abstract**

**Background**: Bibliometric and Scientometric analyses offer invaluable perspectives on the complex research terrain and collaborative dynamics spanning diverse academic disciplines. **Purpose**: This paper presents pyBibX, a python library devised to conduct comprehensive bibliometric and scientometric analyses on raw data files sourced from Scopus, Web of Science, and PubMed, seamlessly integrating state-of-the-art Artificial Intelligence (AI) capabilities into its core functionality. **Methods**: The library executes a comprehensive Exploratory Data Analysis (EDA), presenting outcomes via visually appealing graphical illustrations. Network capabilities have been deftly integrated, encompassing Citation, Collaboration, and Similarity Analysis. Furthermore, the library incorporates AI capabilities, including Embedding vectors, Topic Modeling, Text Summarization, and other general Natural Language Processing tasks, employing models such as Sentence-BERT, BerTopic, BERT, chatGPT, and PEGASUS. **Findings**: As a demonstration, we have analyzed 184 documents associated with “multiple-criteria decision analysis” published between 1984 and 2023. The EDA emphasized a growing fascination with decision-making and fuzzy logic methodologies. Next, Network Analysis further accentuated the significance of central authors and intra-continental collaboration, identifying Canada and China as crucial collaboration hubs. Finally, AI Analysis distinguished two primary topics and chatGPT’s preeminence in Text Summarization. It also proved to be an indispensable instrument for interpreting results, as our library enables researchers to pose inquiries to chatGPT regarding bibliometric outcomes. Even so, data homogeneity remains a daunting challenge due to database inconsistencies. **Significance**: PyBibX is the first application integrating cutting-edge AI capabilities for analyzing scientific publications, enabling researchers to examine and interpret these outcomes more effectively. pyBibX is freely available at https://bit.ly/442wD5z

## 1. Introduction

Bibliometric and scientometric analysis can analyze and quantify the characteristics of a particular literature collection. Both methods use mathematical and statistical techniques; however, the bibliometric analysis focuses on books and forms of written communication (OTLET, 1934; PRITCHARD, 1969; FAIRTHORNE, 1969), and the other focus purely on collections of scientific research (SENGUPTA, 1992). To a broad extent, both methods can overlap concerning the object of study and analysis techniques; therefore, we can assume that both terms may be used interchangeably depending on the study.

Considering a particular collection, some quantitative measures can be taken, such as the total number of documents, the average citation counts, the average journal impact factor, the average h-index, the average publication per author, the total number of research institutions, and many others. Also, patterns and trends in the scientific literature can be identified through Citation Networks, Co-citation Networks, and Artificial Intelligence analysis (SRISAWAD & LERTSITTIPHAN, 2021; BIRCAN & SALAH, 2022).

The analysis output may be used to evaluate the productivity and impact of researchers, journals, and research institutions and identify the strength and weaknesses of research areas. For a simple example, these outputs could help a decision-maker to identify the most influential researchers or institutions, trends, or emerging trends in a particular field of study and patterns of collaboration between researchers and potential collaborators in a specific area. A decision-maker with these pieces of information can make funding decisions and develop research policies and strategies.

Though undeniably, as with any other technique, although the application of the methods can be promising at first glance, they have limitations and potential problems. Some main issues are a) Misuse – the quantitative measures can be over-interpreted, leading to inaccurate evaluations. b) Citation Data – the analysis often relies on citation data, which can be biased and incomplete. c) Selection Bias – the analyzed collection may not represent the entire body of research in a given field. d) Citation Bias – some researchers may attempt to manipulate the quantitative measures by self-citing their work or engaging in citation cartels. This practice may provide false information about the quality or significance of the research (HUG, 2003; AHMED et al., 2012; MIRIAN et al., 2021).

All the mentioned problems are severe, but the collected benefits that this type of analysis offers can exceed the adverse effects caused by these problems. Bearing that in mind, it is natural that many research fronts were developed through the years to spread and facilitate the mindful use of bibliometrics. These research fronts led to two sequential events, which were responsible for our modern scenario of literature research, the creation of scientific databases and the development of software that could deal with this detailed scientific literature data (BELLIS, 2009; BOURNE et al., 2003).

Various databases come into existence; however, for complex reasons, some have stood out, for example, the Web of Science (WoS), Scopus, and PubMed. WoS is a paid-access platform operated by Clarivirate that provides access to an extensive collection of scientific literature and covers a wide range of disciplines, including science, social sciences, and humanities. Scopus is a paid-access platform operated by Elsevier and also covers a wide range of disciplines, including science, technology, medicine, social sciences, and humanities. Finally, PubMed is a free database operated by the National Institutes of Health (NIH) covering biomedical literature worldwide (FALAGAS et al., 2008).

The databases mentioned above provide the possibility to search and discover relevant articles, books, conference proceedings, and other materials through the use o query or search string. The results are compiled in a single file containing pieces of information (metadata), such as authors' names, institutions, abstracts, and many others for each document. Of course, these metadata may conflict between the same document for different databases. Also, it is relevant to note that each database may provide unique types of metadata, such as the date and time a document was accepted (KOKOL & VOŠNER, 2018).

Many bibliometric softwares were developed to work with the files generated by the databases, which may contain an abundant number of documents depending on the search string. The main concern of bibliometric softwares is to provide accurate, reliable, and transparent metrics that can be used to evaluate research impact and productivity, requiring careful attention to data quality and methodology.

It is also worth noting that Artificial Intelligence has recently helped advances in numerous fields of knowledge, including Natural Language Processing (NLP), tremendously. These advances could occur thanks to deep learning models, or in other words, neural network models that can learn from large amounts of data. Artificial Intelligence and bibliometrics can, for example, be used in conjunction to identify clusters of research topics (Topic Modeling), summarize relevant concepts and relationships (Text Summarization), and analyze many bibliometric results (NLP tasks using chatGPT). These tasks are possible due to using Embeddings, a technique to represent words, phrases, or other objects in a high-dimensional vector space. These vectors are designed to capture the semantic and contextual meaning of the objects they represent. Transforming words into vectors and then using the same vectors as input to Deep Learning Models make it possible to develop algorithms to understand and generate human-like language (DEVLIN et al., 2019; ZHANG et al., 2020; REIMERS & GUREVYCH, 2019).

Hence, pyBibX, a python library powered with Artificial Intelligence tools, is one more contribution to bibliometric analysis, a vital tool in scientific research, providing insights into research trends, identifying essential researchers, and evaluating the essential impact of scientific publications. However, traditional bibliometric analysis methods are often time-consuming and require manual effort. Using Artificial Intelligence techniques can significantly enhance the accuracy and efficiency of bibliometric analysis. With the exponential growth of scientific publications, there is a need for more sophisticated bibliometric analysis techniques that can cope with the vast amount of data generated. Artificial Intelligence techniques can process large amounts of data quickly, accurately, and objectively, enabling researchers to gain insights that would otherwise be impossible to obtain. The insights gained from the bibliometric analysis can inform research funding decisions, guide strategic planning, and aid in developing policies to promote scientific research. Therefore, creating a tool that performs bibliometric analysis with Artificial Intelligence techniques can have significant benefits for the scientific community and society as a whole.

The article is divided into five sections, including the introduction. In Section 2, we discuss pyBibX's primary objectives and engage in a comparative analysis of its functionalities vis-à-vis other bibliometric software. Section 3 presents the relevant metadata used by our library and the main difficulties in merging information from the different databases. In section 4, we show the library's capabilities concerning Exploratory Data Analysis (EDA), Network Analysis, and application of Artificial Intelligence tools. A discussion about the obtained bibliometric analysis is also made. Finally, section 5 presents the concluding remarks.

## 2. Related Works

This section presents a comparative assessment of pyBibX's unique characteristics alongside other notable bibliometric tools. Through this analytical juxtaposition, we seek to determine the relative effectiveness and merits of pyBibX, thus highlighting the potential advantages of employing pyBibX; we try to cultivate a comprehensive understanding of its role in shaping the future landscape of bibliometric analysis.

There are many bibliometric softwares available for conducting bibliometric analysis and also visualization. Some outstanding examples are shown in Table 01.

**Table 01 – Bibliometric Tools**

| Name | Data Sources | Environment | Version |
|---|---|---|---|
| pyBibX | PubMed, Scopus, WoS | Python Library | 2.9.3 (16-April-2023) |
| Bibexcel<br>(Persson et al., 2009) | WoS<br>(Needs Data Adaptation) | Windows (Excel) | 1.6.4 (2017) |
| Bibliometrix<br>(Aria & Cuccurullo, 2017) | Cochrane, Dimensions, PubMed, Scopus, WoS | R Package | 4.1.2 (07-March-2023) |
| BiblioTools<br>(Grauwin & Jensen, 2011) | Scopus, WoS | Python Library | 3.2 (01-March-2018) |
| CiteSpace<br>(Chen, 2006) | Wos | Java-Based | 6.2.R2 (26-March-2023) |
| CitNetExplorer<br>(Van Eck & Waltman, 2014) | Scopus, WoS | Java-Based | 1.0.0 (10-March-2014) |
| CRExplorer<br>(Thor et al., 2016) | Scopus, WoS | Java-based | 1.9 (16-July-2018) |
| Litstudy<br>(Heldens et al., 2022) | Scopus and Others | Python Library | 1.0.5 (28-March-2023) |
| Metaknowledge<br>(McLevey, 2017) | PubMed, Scopus, WoS, and Others | Python Library | 3.4.1 (03-November-2021) |
| Publish or Perish<br>(Harzing, 2010) | PubMed, Scopus, WoS, and Others | Windows | 8.2.3944 (12-May-2022) |
| Sci2 Tool<br>(Börner et al., 2003) | Scopus, WoS, and Others | Windows, MacOS, Linux | 1.3.0 (24-June-2020) |
| Scientopy<br>(Ruiz-Rosero et al., 2019) | Scopus, WoS | Python Library | 2.1.2 (24-March-2023) |
| SciMat<br>(Cobo et al., 2012) | Scopus, WoS | Windows, MacOS, Linux | 1.1.04 (12-July-2016) |
| Tethne<br>(Peirson et al., 2016) | Scopus, WoS | Python Library | 0.8 (21-June-2016) |
| VOSviewer<br>(Van Eck & Waltman, 2010) | Dimensions, PubMed, Scopus, WoS | Windows, MacOS, Linux | 1.6.19 (23-January-2023) |

Table 01 compares 15 bibliometric analysis tools, showcasing their unique and common features. The table includes, in addition to the Name column, the following columns: Data Sources, Environment, and Version.

a) Data Sources: Each tool or library utilizes specific data sources, the most prevalent being PubMed, Scopus, and WoS. Bibliometrix, LitStudy, Metaknowledge, Publish or Perish, and VOSviewer extends their compatibility to include other databases like PubMed, Dimensions, Cochrane, and others. Notably, Bibexcel necessitates adjustments to function with the WoS database.

b) Environment: This column highlights the environment or platform on which each tool functions. Numerous tools, such as pyBibX and BiblioTools, operate as Python libraries, while others, like CiteSpace and CitNetExplorer, are Java-based applications. Bibexcel and Publish or Perish are designed only for Windows, and Bibliometrix is available as an R package. Meanwhile, Sci2 Tool, SciMat, and VOSviewer provide multiplatform support for Windows, macOS, and Linux. The choice of environment ultimately depends on user preferences, accessibility, and specific project requirements.

c) Version: This column presents each tool's version number and release year. Recently updated tools, such as pyBibX (2.9.3, 16-April-2023), Bibliometrix (4.1.2, 07-March-2023), and CiteSpace (6.2.R2, 26-March-2023), are more likely to include the latest features and enhancements. On the other hand, older tools like Bibexcel (1.6.4, 2017) and Tethne (0.8, 21-June-2016) might not offer the most recent improvements. Version numbers and updates are crucial as they indicate that developers are actively releasing updates to refine features, fix bugs, or introduce new functionalities, ensuring their tools remain compatible with modern bibliometric analysis techniques.

The bibliometric tools, as presented in Table 01, display a range of similarities and discrepancies across the Data Sources, Environment, and Version columns. Despite the differences, these bibliometric tools cater to diverse research requirements and preferences, providing researchers with a wide array of bibliometric analysis and visualization options to suit their needs.

Table 02 offers an in-depth comparison of the bibliometric tools featured in Table 01, emphasizing their EDA, Network Analysis, and Artificial Intelligence capabilities. EDA capabilities encompass the tools' capacity for preliminary data analysis, pattern and trend identification, and data visualization, empowering researchers to uncover insights and formulate hypotheses. Network Analysis capabilities indicate the tools' proficiency in examining and visualizing intricate relationships among entities such as authors, citations, and keywords, thereby aiding researchers in deciphering data structure and dynamics. Artificial Intelligence capabilities utilize Deep Learning techniques, like Topic Modeling, Embedding vectors, Text Summarization, and General NLP tasks, to augment the tools' effectiveness.

**Table 02 – Key Features**

| **Name** | Data Manipulation | EDA | Wordcloud | N-Gram | Projection | Evolution Plot | Sankey Diagram | Treemap | Bar Plot | Citation Analysis | Collaboration Analysis | Similarity Analysis | World Collab. Analysis | Topic Modeling | Embeddings | Text Summarization | chatGPT |
|---|---|---|---|---|---|---|---|---|---|---|---|---|---|---|---|---|---|
| pyBibX | *X* | *X* | *X* | *X* | *X* | *X* | *X* | *X* | *X* | *X* | *X* | *X* | *X* | *X* | *X* | *X* | *X* |
| Bibexcel | | *X* | | | | | | | | *X* | | *X* | | | | | |
| Bibliometrix | *X* | *X* | *X* | | *X* | *X* | *X* | | *X* | *X* | *X* | *X* | *X* | | | | |
| BiblioTools | | *X* | | | | | | | | *X* | | *X* | | | | | |
| CiteSpace | | | | | | | | | | *X* | *X* | *X* | *X* | | | | |
| CitNetExplorer | | | | | | | | | | *X* | *X* | *X* | *X* | | | | |
| CRExplorer | | | | | | | | | | *X* | *X* | *X* | | | | | |
| Litstudy | | *X* | | | | | | | *X* | *X* | *X* | *X* | *X* | *X* | | | |
| Metaknowledge | | | | | | | | | | *X* | *X* | *X* | *X* | | | | |
| Publish or Perish | | *X* | | | | | | | | | | | | | | | |
| Sci2 Tool | | *X* | | | | | | | | *X* | *X* | *X* | *X* | | | | |
| Scientopy | *X* | *X* | *X* | | | *X* | | | *X* | *X* | *X* | *X* | *X* | *X* | | | |
| SciMat | | *X* | | | | | | | *X* | *X* | *X* | *X* | *X* | | | | |
| Tethne | | | | | | | | | | *X* | *X* | *X* | *X* | *X* | | | |
| VOSviewer | | | | | | | | | | *X* | *X* | *X* | *X* | | | | |

In Table 02, the most common features these tools share include Citation Analysis, Collaboration Analysis, and World Collaboration Analysis, with many tools supporting Similarity Analysis and Topic Modeling. Both pyBibX and Scientopy offer a wide range of features, covering most of the functionalities in Table 02. CiteSpace, CitNetExplorer, CRExplorer, Metaknowledge, Tethne, and VOSviewer primarily focus on citation-related analysis, with few additional features. Bibexcel, BiblioTools, Litstudy, Sci2 Tool, and SciMat concentrate on data manipulation and/or EDA, with some also providing citation, collaboration, and topic analysis. R-Package Tool: Bibliometrix, the only R package in the list, offers a variety of features, including data manipulation, EDA, and various types of analysis. However, pyBibX stands out with its extensive feature set, including unique features such as Embedding, Text Summarization, and chatGPT integration. On the other hand, Publish or Perish is notable for its minimalistic approach, focusing primarily on EDA.

Upon examining both tables, we can discern several insights and patterns: pyBibX, BiblioTools, Litstudy, Metaknowledge, Scientopy, and Tethne are python libraries with varying functionalities. pyBibX and Scientopy are the most feature-rich among these, providing data manipulation, EDA, and various visualization and analysis techniques, making them suitable for researchers looking for extensive functionality within a single tool. Litstudy and SciMat also offer a decent range of features, while BiblioTools and Tethne are more specialized in citation and collaboration analysis. Java-Based Applications like CiteSpace, CitNetExplorer, and CRExplorer primarily focus on citation, collaboration, and similarity analyses. They share similar functionalities but are not as comprehensive as some python libraries. Bibexcel and Publish or Perish are Windows-based tools with unique features. Bibexcel requires data adaptation for WoS plain text and relies on Excel, while Publish or Perish supports a broader range of data sources. However, both tools have limited functionalities compared to others in the list. Bibliometrix is the only R package in the list. It offers a wide array of features, including data manipulation, EDA, visualization, and various types of analysis, making it a versatile option for researchers familiar with R. Also, Bibliometrix, Litstudy, Metaknowledge, and Publish or Perish extend their support to various data sources, making them more versatile for researchers working with different databases. Sci2 Tool, SciMat, and VOSviewer are compatible across Windows, macOS, and Linux platforms, each offering a different range of features. Sci2 Tool and SciMat cover multiple analysis and visualization techniques, while VOSviewer focuses on visualization and lacks data manipulation and EDA capabilities.

Integrating pyBibX and chatGPT in bibliometric analysis offers a groundbreaking approach that enhances the field of bibliometrics as a whole. The fusion of chatGPT with pyBibX streamlines the process of interpreting and understanding complex bibliometric results. ChatGPT's conversational features enable users to interact seamlessly with the analysis software, asking questions and obtaining pertinent answers. This exchange facilitates the clarification of insights and patterns, allowing researchers to readily comprehend essential information and recognize trends without engaging in laborious manual data exploration.

Moreover, the integration of chatGPT with pyBibX fosters interdisciplinary research by bridging the gap between bibliometrics and NLP. This synergy promotes the development of novel methods and techniques that can further advance both fields. As a result, researchers can benefit from more robust and efficient bibliometric tools, enabling them to tackle increasingly complex research questions. This powerful combination is poised to reshape how researchers approach bibliometric analysis, driving innovation and unlocking new insights.

## 3. Scientific Databases and Metadada Utilization

Three of the most commonly used scientific databases are WoS, Scopus, and PubMed. While these databases share similarities in their coverage of scientific literature, they also have unique features that set them apart. Researchers working in a specific field, such as medicine, may find PubMed the most relevant, as it covers biomedical literature in-depth. On the other hand, researchers working in interdisciplinary fields may find WoS or Scopus more valuable, as they cover a broad range of subjects.

Regarding the document's metadata provided by these scientific databases, all of them are a set of structured information that describes the content and context of each document. The structured information is not the same between the databases. Differences may arise due to various factors, including indexing methods, errors or omissions, distinct document versions, and the database scope and focus. Therefore, chances are that a document indexed in more than one database may present divergent metadata. For example, the total number of citations may present different total counts. Table 03 shows the metadata items and the technique(s) each metadata allows. It's worth mentioning that all metadata items can be obtained from all three databases except for "*References*" and "*Times Cited*", which are not included in PubMed.

**Table 03 – Databases Metadata**

| Metadata | Definition | Exploratory Data Analysis | Network Analysis | Artificial Intelligence |
|---|---|---|---|---|
| Abstract | A summary of the document's main points and findings gives readers a quick overview of the document's purpose and conclusions. | Evolution Plot, N-Grams, Projection, Wordcloud | | Embeddings, Text Summarization, Topic Modeling |
| Author(s) | The authors who contributed to the document. | Authors Productivity Plot, Bar Plot, EDA Report, Sankey Diagram, | Collaboration Analysis | |
| Author(s) Affiliation | The name, address, and country of the organization(s) where the authors are affiliated or employed. | EDA Report, Bar Plot, Sankey Diagram | Collaboration Analysis, World Map Collaboration Analysis | |
| Author(s) Keywords | A list of keywords or phrases chosen by the authors that summarize the main topics or themes of the document. | Bar Plot, EDA Report, Evolution Plot N-Grams, Projection, Sankey Diagram, Wordcloud | Adjacency Analysis | Embeddings |
| Document Title | The title of the document | Evolution Plot, N-Grams, Projection, Wordcloud | | Embeddings |
| Document Type | The document type (original research article, review article, etc.). | EDA Report | | |
| Journal | The name of the journal in which the document is published. | Bar Plot, Evolution Plot, Sankey Diagram | | |
| Keywords Plus | Additional keywords or phrases that supplement the author-chosen keywords and provide more context for the document. | Bar Plot, EDA Report, Evolution Plot, N-Grams, Projection, Sankey Diagram, Wordcloud | Adjacency Analysis | Embeddings |
| Language | The language in which the article is written. | Bar Plot, EDA Report, Sankey Diagram | | |
| *References* (Not in PubMed) | A list of the references cited in the document. | EDA Report | Similarity Analysis | |
| *Times Cited* (Not in PubMed) | The number of times other researchers have cited the document. | Bar Plot, EDA Report | Citation Analysis, Similarity Analysis | |

Table 03 indicates the metadata used by our library, including their definitions and the technique that uses the specific metadata. The techniques are separated into three groups, EDA, Network Analysis, and Artificial Intelligence.

The EDA in Table 1 refers to the various techniques and methods that can be applied to analyze the metadata of a document to uncover patterns, relationships, and trends. The techniques that are listed in Table 1 include:

1) Author's Productivity Plot: A graphical representation of an author's productivity over time. It helps track an author's output and identify research patterns.
2) Bar Plot: A common type of plot used to display categorical data and is particularly useful for comparing two variables. Our library offers many bar plot options, which will be discussed in section 3. Essentially each option provides a unique perspective on the research metadata and can be used to gain insights into the research landscape, such as the most influential authors, the most cited documents, and the most common keywords used in the field. By visualizing these data points on a bar plot, researchers can better understand the trends and patterns in the metadata and use this information to inform their research endeavors.
3) EDA Report: A summary report that provides descriptive statistics and visualizations of the metadata. It is an essential step in data analysis to gain insights into the data, identify patterns and relationships, and verify assumptions. The report provides the timespan of the research, the total number of countries and institutions, and the total number of sources and references. The report also includes information on the languages used in the documents, the total number of documents, the document types, and the average number of documents per author, institution, source, and year. Additionally, the report provides information on the number of authors, author keywords, and collaborations. Finally, the report includes information on the highest local h-Index and the total number of citations, with averages for citations per author, institution, document, and source.
4) Evolution Plot: This is a graphical representation of the change in frequency of a term or topic over time. It is a valuable tool for visualizing trends and patterns in data and can be used to identify changes in behavior or patterns over time.
5) N-Grams: A technique for extracting patterns of words from a text by dividing the text into contiguous sequences of $n$ words.
6) Projection: A method to transform high-dimensional data into a lower-dimensional space while preserving the structure and relationships between data points. Data projection aims to reduce the complexity of high-dimensional data and visualize it more efficiently and interpretably. Two algorithms may be used, Uniform Manifold Approximation and Projection (UMAP), developed by McInnes et al. (2018), or Truncated Singular Value Decomposition (TSVD).
7) Sankey Diagram: This diagram shows the flow of documents between any combination of seven elements (authors, countries, institutions, sources, author's keywords, keywords plus or languages). In a Sankey Diagram, the width of the lines represents the flow of research production, and the nodes represent the different elements.
8) Wordcloud: This is a graphical representation of the most frequently occurring words in a text, with the size of each word reflecting its frequency. They are helpful for quickly identifying the most common themes or topics in a large corpus of text data. It is important to note that wordclouds have limitations, as they do not provide a detailed analysis of the content or context of the text data. Therefore, they should be used with other text analysis techniques for a more comprehensive understanding of the data.

The Network Analysis column in Table 1 refers to the techniques used to analyze the relationships and interactions between different elements in the metadata, as they can be useful in understanding the underlying structure and dynamics of all research. These techniques include:

1) Adjacency Analysis: It is used to identify the proximity and connections between different elements in the metadata and visualize the resulting network structure.
2) Collaboration Analysis: It is used to identify the co-occurrence of authors or institutions in the metadata and visualize the collaboration patterns between them.
3) World Map Collaboration Analysis: It is used to map the locations of authors or institutions in the metadata and to visualize the patterns of collaboration between them on a global scale.

The Artificial Intelligence column in Table 1 refers to the techniques used to apply Deep Learning algorithms to the metadata to generate insights and reveal patterns. These techniques include:

1) Embeddings: This technique can represent words or phrases as vectors in a high-dimensional space, which can be used to identify similarities and relationships between different elements in the metadata.

2) Text Summarization: This is a technique used to generate a summary of the text data in the metadata, which can be useful in quickly understanding the main points and themes of the document. Abstractive and Extractive Summarizations can be applied.
3) Topic Modeling: This is a technique used to identify the underlying topics or themes in the metadata, which can be useful in understanding the structure and content of the document.
4) General NLP Tasks: This feature uses chatGPT to analyze bibliometric results. The results that can be analyzed are EDA Report, WordCloud, N-Grams, Evolution Plot, Sankey Diagram, Authors Productivity Plot, Bar Plots, Citation Analysis, Collaboration Analysis, Similarity Analysis, and World Map Collaboration Analysis. The researcher can input any question to help understand the bibliometric results.

The final consideration about the metadata in the databases, as mentioned earlier, regards merging the information of two or more databases. Merging metadata from different sources such as Scopus, WoS, and PubMed can be challenging due to several issues. First, these databases may use different formats and standards for their metadata, which can create inconsistencies and errors when merging the data. Second, the databases may have different coverage and selection criteria, resulting in variations in the data available for analysis. Third, the databases may have different levels of accuracy and completeness, which can affect the quality of the merged data. Finally, the metadata may contain errors, duplicates, or missing values, further complicating the merging process. Although our library permits manually editing the metadata elements, it solves only small-scale problems, as correcting a large number of documents proves to be an unfathomable work.

One potential solution is to follow a specific order and avoid overwriting the reference metadata. This approach involves selecting a database (e.g., PubMed) and using it as a reference for merging the data from the other databases. The newer metadata is then added to the reference data, but only new information not already present in the reference is included. This process avoids overwriting the existing metadata and ensures the merged dataset contains all available information without duplications.

However, one can note that the sequence of this merging process preserves the prior databases, and the merge order may affect the final result. For example, merging a WoS dataset followed by a Scopus dataset will not generate the same result as merging a Scopus dataset followed by a WoS dataset. Starting with either the WoS or Scopus dataset for the metadata merging process reflects the researcher's confidence in the chosen database. Adding data from the other database then complements the existing dataset. If the merging process begins with WoS, Scopus data will add to the metadata, and vice versa.

Notwithstanding, by following this approach, it is still possible to create a high-quality and comprehensive dataset that leverages the strengths of multiple databases while avoiding the challenges associated with merging inconsistent and incomplete metadata.

## 4. pyBibX – Set of Documents

As we aim to showcase the powerful capabilities of our python library, pyBibX, we have opted to employ a search string of limited scope in our demonstration. This decision made it easier for readers to comprehend our highlighted concepts. However, it is important to note that our library can handle far more complex search strings and offers users an extensive range of functionalities to explore. We encourage users to visit the pyBibX homepage at https://pypi.org/project/pyBibX to learn more about the library's features and explore its capabilities.

To demonstrate the library's capabilities, we have searched in the Scopus, WoS, and PubMed databases documents containing in the title the following query - "*multiple-criteria decision analysis*" – published until the 1st of January of 2023. We have downloaded all available information each database can offer for all documents. The obtained files from Scopus (mcda_scopus.bib), WoS (mcda_wos.bib), and PubMed (mcda_pubmed.txt) can be accessed at bit.ly/3XYK3eB.

The search yielded 234 documents in Scopus, 206 in WoS, and 10 in PubMed. However, only 30 documents differed from Scopus in WoS, and only one document differed from the combined search of Scopus and WoS in PubMed. The research following the Scopus, Wos, and PubMed merging order totaled 265 documents.

Our library can also filter documents by various criteria, document type, year, sources, sources according to Bradford's Law (BRADFORD, 1934), country, language, and the presence of an abstract. Bradford's Law is a statistical pattern in bibliographic data where a small number of sources contribute significantly to the literature of a particular research field. Bradford's Law is usually represented as a three-zone model, where the core sources (Zone 1) contribute the most significant number of documents, followed by a relatively minor number of documents from the second zone, and finally, many documents from the third zone.

Then we had one additional step where the 265 documents were filtered, and only documents classified as "Articles", "Conference Papers", "Meeting Abstracts", and "Reviews" were accepted for analysis. There were no restrictions on the years, sources, or countries from which the documents were retrieved. However, the documents had to be written in English and contain an abstract. This filter resulted in 184 documents representing our final dataset that will be analyzed in the following sections.

All 184 documents were given an ID number starting from 0 to 183. Additionally, each author was assigned an ID label "a_#", where # starts with 0 and ends with the total number of elements minus one. Similarly, each source was assigned an ID label "j_#", and the institutions were assigned a label "i_#". Each country with the label "c_#". Each author's keyword was labeled "k_#", and each keyword plus was labeled "p_#". Furthermore, each reference cited in each document receives an "r_#" label if available; the only exception is when the reference is one of the 184 documents, so its original label remains. This labeling system helps to identify and visualize different metadata elements in a less convoluted manner.

The following sections include EDA, Network Analysis, and Artificial Intelligence techniques to analyze the metadata elements further and generate insights. These steps can be accessed through bit.ly/3mcP2eb, which provides a user-friendly online interface for conducting these analyses and exploring the data in greater detail.

In addition to using the given link, users can install the library directly through their preferred Integrated Development Environment (IDE) using the command: *pip install pyBibX*. Once installed, the pyBibX package can be imported into a python script, allowing users to access its functions and perform bibliometric analyses on their own datasets.

### 4.1 pyBibX – Exploratory Data Analysis

In this section, we will perform EDA on the merged dataset. This essential step in data analysis helps us understand the data's underlying structure, patterns, and relationships. Also, this step provides a comprehensive overview of the merged dataset and enables us to identify key features and trends that may inform further analysis. Table 04 shows the preliminary statistical report.

**Table 04 – Merged Metadata Report**

| Main Information | Results |
|---|---|
| Timespan | 1984-2023 |
| Total Number of Countries | 45 |
| Total Number of Institutions | 231 |
| Total Number of Sources | 121 |
| Total Number of References | 7926 |
| Total Number of Languages | 1 |
| --English (# of docs) | 184 |
| | |
| Total Number of Documents | 184 |
| --Article | 184 |
| Average Documents per Author | 1.2 |
| Average Documents per Institution | 2.64 |
| Average Documents per Source | 1.52 |
| Average Documents per Year | 7.36 |
| | |
| Total Number of Authors | 495 |
| Total Number of Authors' Keywords | 587 |
| Total Number of Authors' Keywords Plus | 1434 |
| Total Single-Authored Documents | 43 |
| Total Multi-Authored Documents | 141 |
| Average Collaboration Index | 3.21 |
| Max h-Index | 21 |
| | |
| Total Number of Citations | 5674 |
| Average Citations per Author | 11.46 |
| Average Citations per Institution | 24.56 |
| Average Citations per Document | 30.84 |
| Average Citations per Source | 46.89 |

Table 04 provides various statistics related to the collection of documents comprehending the timespan of 1984-2023. The documents were written in English and were sourced from 121 different journals. The total number of countries represented was 45, and the number of institutions involved in producing the documents was 231. These documents consist entirely of articles and have an average of 1.2 documents per author, 2.64 documents per institution, and 1.52 documents per source. Over the entire period, there were 495 authors and 587 authors' keywords, with 1434 authors' keywords plus. Of the 184 documents, 43 were single-authored, and 141 were multi-authored, with an average collaboration index of 3.21 and a maximum h-index of 21. The documents received 5674 citations, with an average of 11.46 citations per author, 24.56 citations per institution, 30.84 citations per document, and 46.89 citations per source.

Several conclusions can be drawn from the information presented in Table 2. Here are a few possible ones: a) The collection of 184 documents is relatively small, with an average of only 7.36 documents per year over the 40-year timespan covered. b) The average collaboration index of 3.21 indicates that the collection of documents results from multiple authors working together rather than single-authored works. c) The relatively high average citations per source (46.89) suggests that the documents in the collection are well-referenced and draw from a diverse range of sources. d) The maximum h-index of 21 indicates at least one highly cited author in the collection. Still, the average number of citations per author (11.46) suggests this is not typical across all authors. e) The average citations per institution (24.56) suggest that some institutions may be more heavily represented in the collection than others.

While the conclusions drawn from Table 04 provide some valuable insights, there are several areas where a more profound analysis may be warranted. For example, while Table 2 provides some information on collaboration and citation metrics, a more detailed analysis could explore the nature and extent of collaborations within the collection and the patterns of citation and reference across authors, institutions, and sources.

Continuing with the EDA, one way to gain further insights about the documents would be to conduct a text analysis using wordclouds and n-grams. By generating wordclouds and n-grams (n is a positive integer greater than 1) from the abstracts of the documents, it may be possible to identify common themes, topics, and terminology used by the authors. Figure 01 shows the results of both exploratory analyses, a wordcloud, and a 3-Gram (a n-gram with 3 words).

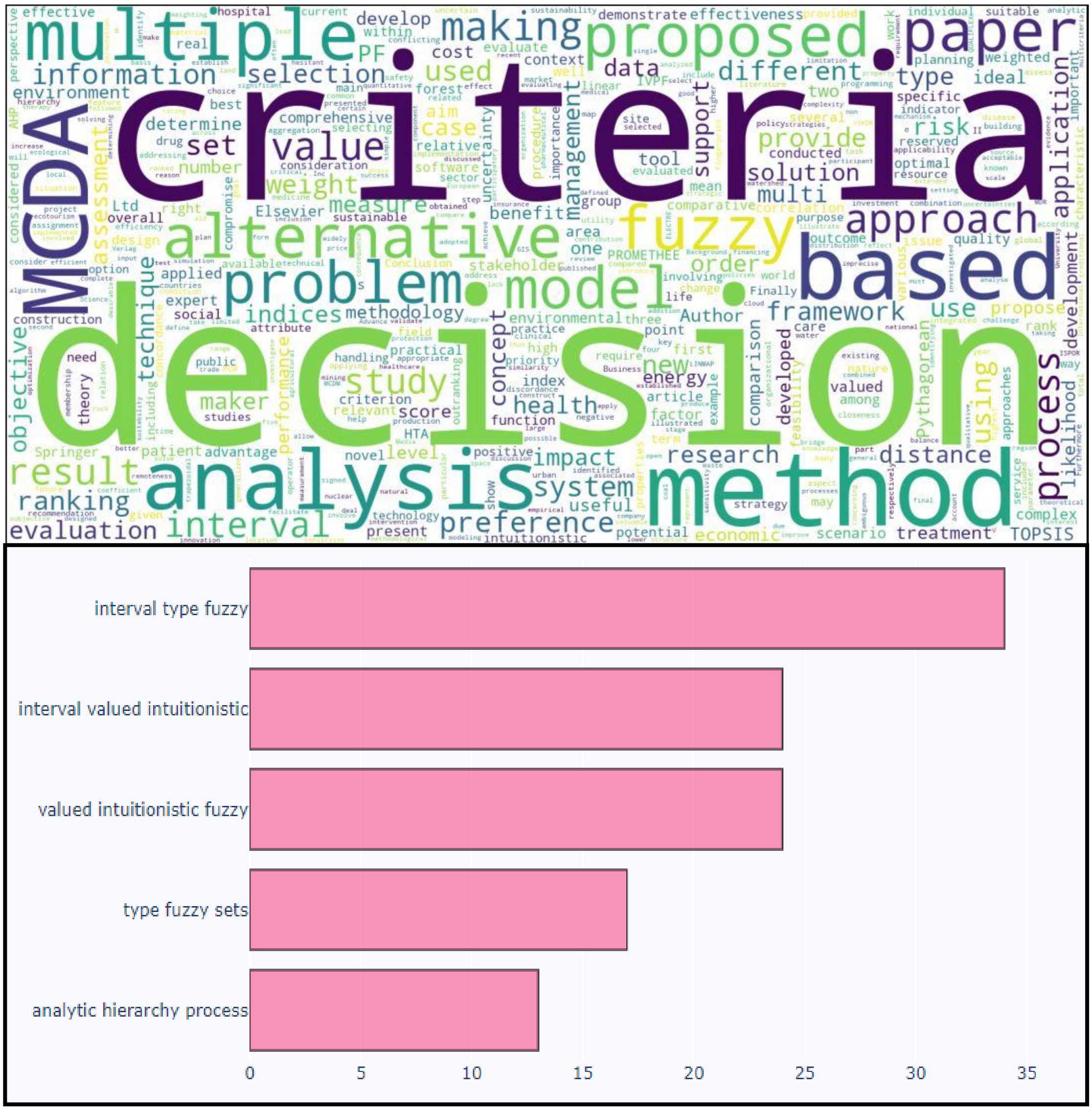

**Figure 01- Wordcloud and 3-Gram**

Based on the highlighted words in the wordcloud (Figure 01 upper graph) generated from the abstracts of the documents, it seems that the collection's primary focus is on decision-making, particularly in the context of Multiple Criteria Decision Analysis (MCDA) and fuzzy logic methods. The prominence of these terms in the wordcloud suggests that the collection of documents may benefit researchers and practitioners interested in decision-making processes and techniques. Further analysis of the collection could explore the specific applications, challenges, and innovations related to MCDA and fuzzy logic methods.

The 3-gram analysis (Figure 01 lower graph) provided insights into the specific concepts and techniques discussed in the documents. However, the documents abstracts were first normalized by removing English stop words and the following custom stop words specific to this field of knowledge: "analysis", "criteria", "decision", "elsevier", "making", "mcda", "multiple", "reserved", and "rigths". After the normalization, the most frequent 3-gram was "interval-type-fuzzy", with a frequency of 34, indicating that this concept is a critical theme in the collection of documents. The 3-grams "interval-valued-intuitionistic" and "valued-intuitionistic-fuzzy" both appeared with a frequency of 24, highlighting the importance of these concepts.

The 3-gram “type-fuzzy-sets” appeared 17 times, indicating that fuzzy sets and related techniques are also relevant topics. Interestingly, the 3-gram “analytic-hierarchy-process” appeared 13 times, confirming that the Analytic Hierarchy Process (AHP) is a well know method for multi-criteria decision-making (MAZUREK et al. 2020; MAZUREK et al. 2021; PEREIRA & COSTA, 2015; FLORIANO et al. 2022). In our library, this graph is interactive, and hovering the cursor over it gives more information about the results.

After exploring the dataset with techniques such as wordclouds and n-grams, the next step in EDA is to transform the abstracts into numerical representations. The abstracts of the 184 documents can be transformed into a Term Frequency-Inverse Document Frequency (TF-IDF) matrix or Embeddings. Once the documents are represented this way, dimension reduction techniques such as UMAP or TSVD can be applied to reduce the dimensionality to two dimensions for visualization purposes. Document clustering based on dimension reduction techniques is widely used to analyze texts; by reducing the dimensionality of the data, these techniques can reveal the underlying structure of the text data, making it easier to explore and interpret (RAMKUMAR & POORNA, 2016). The final step is to group the documents into clusters using the k-means algorithm (McQUEEN, 1967) to group the documents in k natural groups based on a distance measure and plot the results, with each cluster highlighted by a different color. By visualizing the clusters, researchers can quickly identify patterns and relationships within the data and use this information to make more informed decisions about the next steps in their analysis and help researchers to uncover new insights and research directions.

The combination of UMAP and Embeddings is commonly more indicated for this approach because UMAP is a more advanced dimension reduction technique than TSVD and is better suited to preserving the global structure of the data (JEON et al. 2022). Additionally, UMAP is more effective at handling high-dimensional data with complex structures, which is common in NLP. On the other hand, TSVD and TF-IDF are often favored over UMAP and Embeddings for their speed and efficiency, as both techniques are relatively quick and can handle large amounts of text data (DESSI et al. 2021). However, it is essential to note that the choice of technique should ultimately depend on the specific characteristics of the analyzed text data. Figure 02 shows the projection of the 184 document abstracts using the UMAP and Embeddings combination.

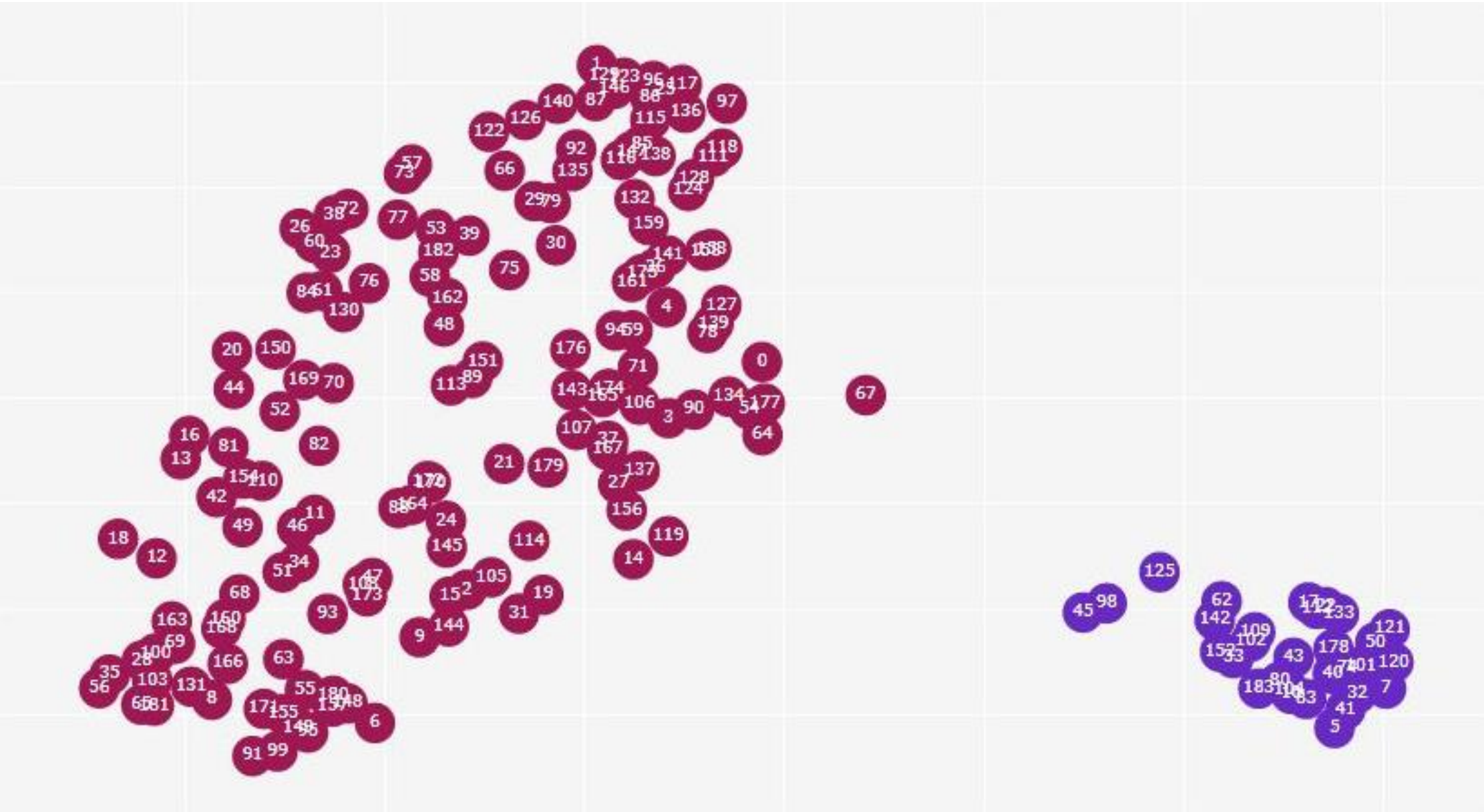

**Figure 02- Documents Projection - UMAP and Embeddings**

In Figure 02, we arbitrarily selected 2 clusters to represent the transformed abstracts visually. This graph is an interactivity plot, and hovering over a node reveals additional document ID, cluster number, and complete document citation information. The projection allowed us to identify a highly homogeneous cluster (purple) distinct from the other cluster, indicating that the documents in this cluster may share a common theme or topic. Further investigation into this cluster may provide insights into this theme or topic and potentially reveal new avenues for research.

Identifying the themes or topics can be further investigated using various techniques. One way is to create an evolution plot of the keywords used in the abstracts over the years, and this can give insight into how the topic has evolved and what subtopics are emerging. By looking at how the frequency and usage of certain words change over time, researchers can gain insights into the shifting focus of the field and the emergence of new topics. Figure 03 shows the evolution plot of the keywords plus from 2018 to 2021.

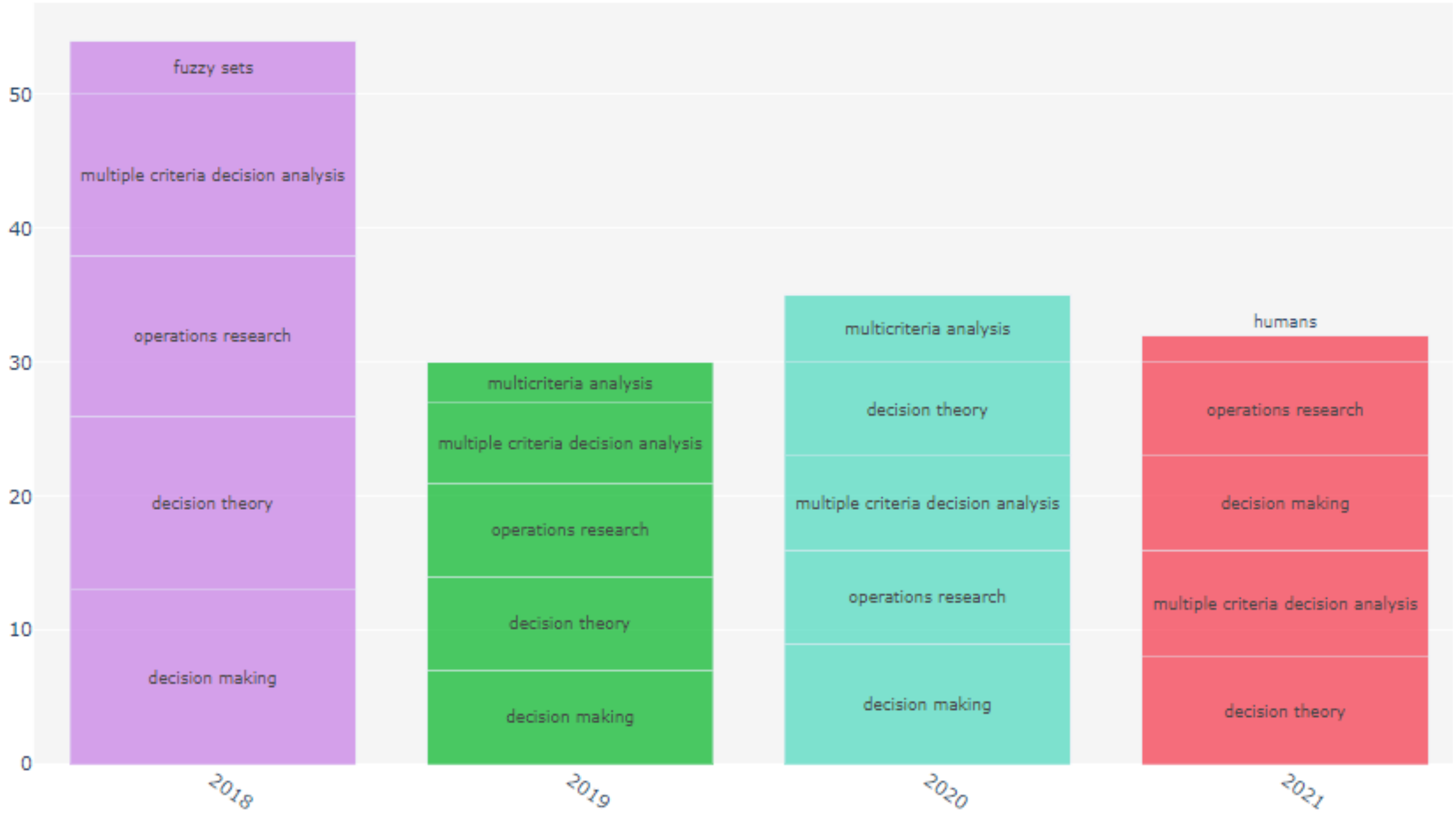

**Figure 03- Evolution Plot 2018 – 2021 Keywords Plus**

Figure 03 revealed some interesting trends in the research topics. The most commonly occurring keywords across all four years were: “multiple criteria decision analysis”, “decision theory”, “operations research”, and “decision making”. These keywords suggest that research in decision-making and analysis is ongoing and active. In particular, the prevalence of “multiple criteria decision analysis” indicates a continued focus on this decision-making approach. Another interesting finding was the appearance of “humans” in the 2021 keywords, which suggests a growing interest in human factors in decision-making.

This technique provides a supportive visualization of the changes in research focus over time. The interactive nature of the graph allows for easy exploration of specific keyword frequencies, and the temporal feature is an essential factor. However, it should be noted that Topic Modeling can also extract deeper insights and may provide a more nuanced understanding of the underlying themes and topics.

Allied with the evolution plot, we can also use a treemap to aid our understanding of the dominant central theme of all 184 documents. By applying this technique to the entire set of keywords plus across all 184 documents, we can quickly identify the most frequently occurring keywords, which may indicate significant themes and topics. Figure 04 shows the results for the top 15 keywords plus.

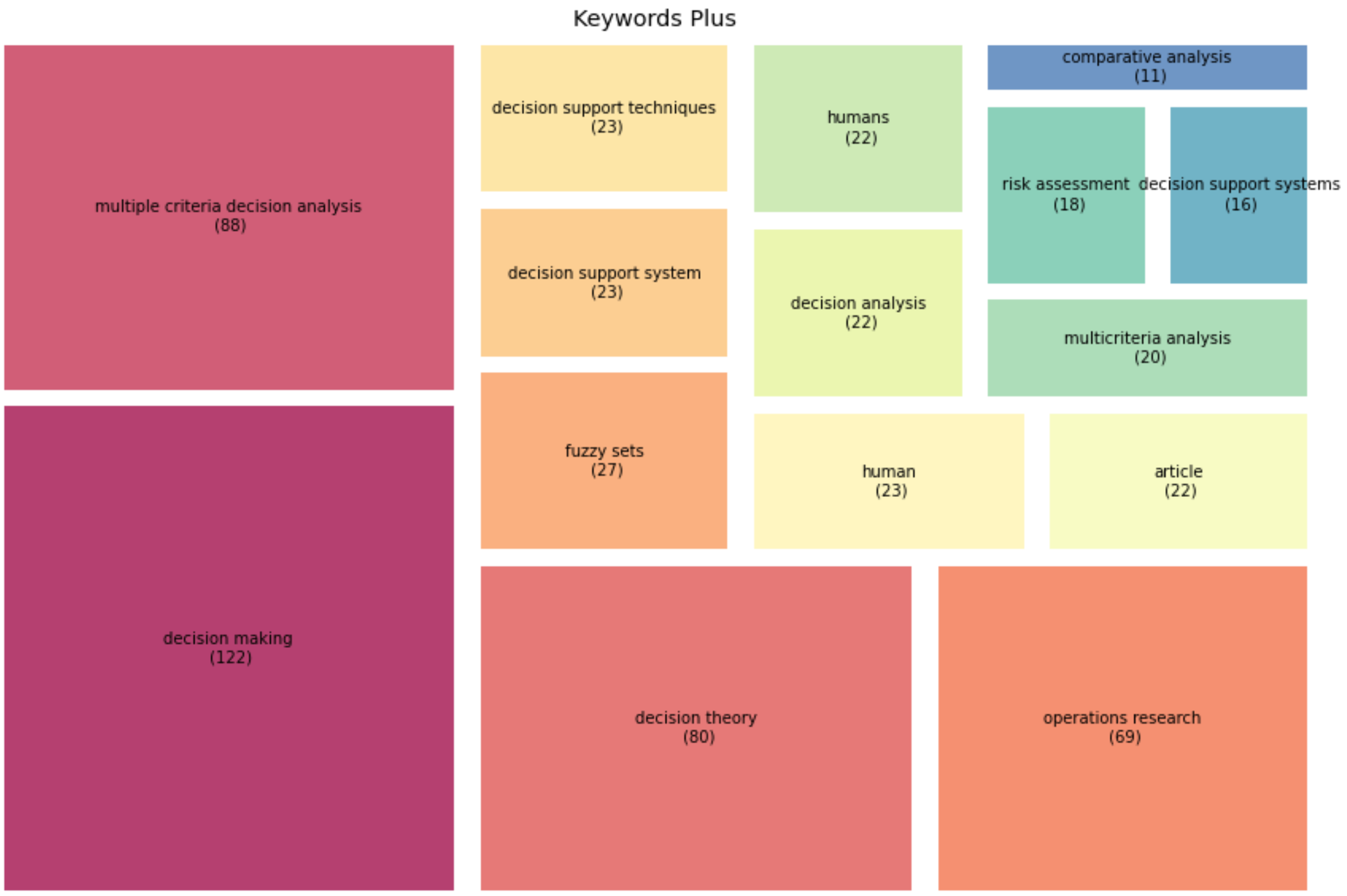

**Figure 04- Treemap Top 15 Keywords Plus**

Figure 04 shows the frequency of keywords plus in the 184 documents. The most common ones s are "decision making", with a frequency of 122, followed by "multiple criteria decision analysis", with 88 occurrences, and "decision theory", with 80 occurrences. "Operations research" and "fuzzy sets" have a frequency of 69 and 27, respectively. The high frequency of the keyword "operations research" in the documents suggests that this discipline is relevant to decision-making and MCDA. The lower frequency of the keyword "fuzzy sets" in the documents suggests that it is a less popular approach when compared to "operations research". However, it still indicates that "fuzzy sets" are a relevant topic in this context and could be further investigated. Again it's interesting to note that the keyword human has a frequency of 23, and humans 22, suggesting that there is a focus on human-centric approaches in decision-making processes.

As mentioned, a Sankey Diagram shows the flow of documents between any combination of seven elements: authors, countries, institutions, sources, author's keywords, keywords plus, or languages. To continue our analysis, we can show the flow between the authors and the keywords plus, and Figure 05 shows the top 10 flows.

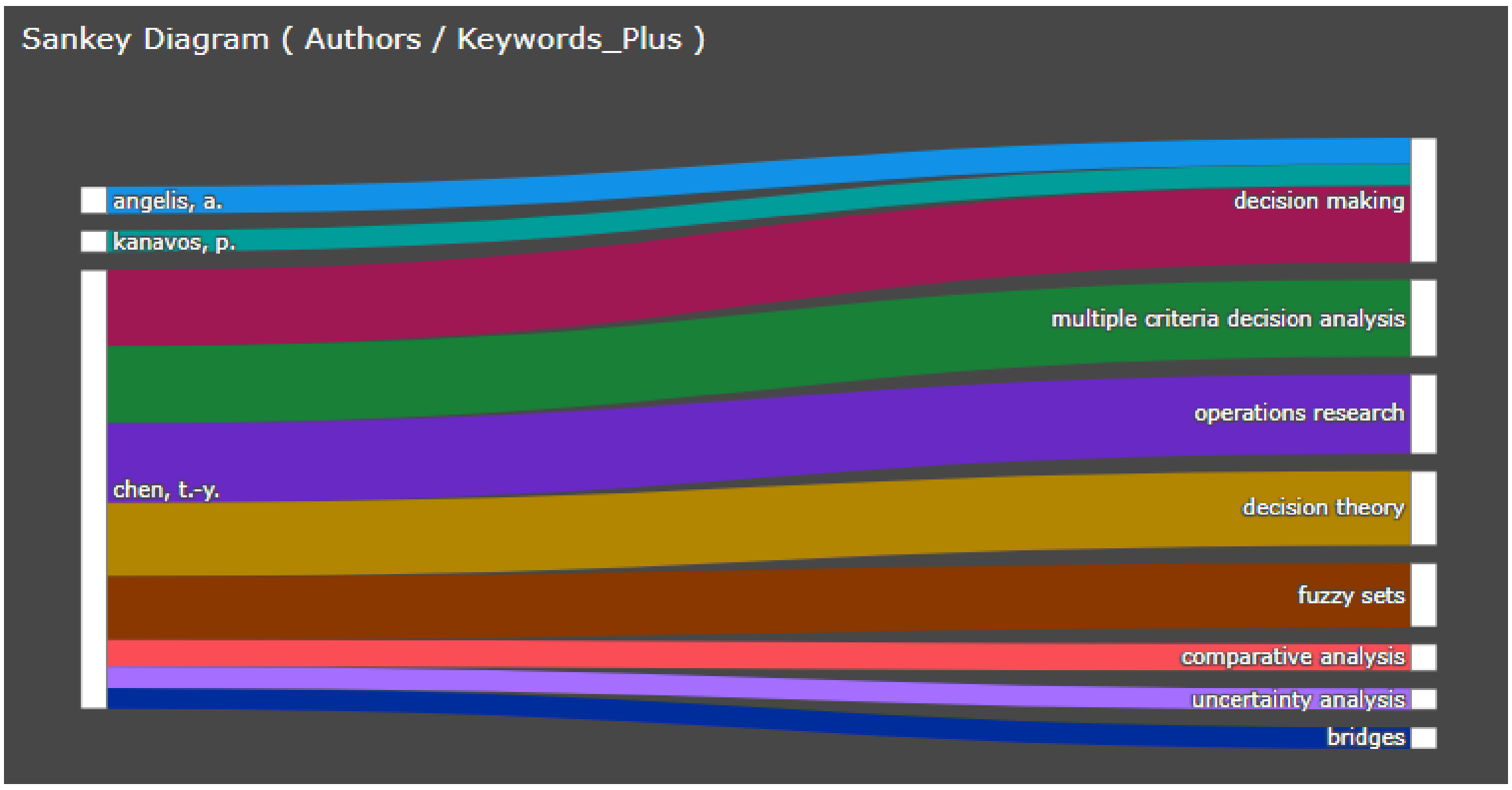

**Figure 05- Sankey Diagram Authors -Keywords Plus**

Figure 05 shows the top 10 flows between authors and the keywords plus Angelis A., and Kanavos, P. with one flow and Chen, T.-Y. with eight. The keywords plus are primarily consistent with the themes identified in the treemap (Figure 04), including “decision making”, “multiple criteria decision analysis”, “operations research”, “decision theory”, and “fuzzy sets”. Remarkably, Chen, T.-Y. appears to be the most prolific author, investigating a wide range of topics. The results suggest potential collaboration possibilities between the authors’ shared interests, creating more contributions to the field and exploring research opportunities.

To confirm that Chen, T.-Y. is a prolific author, we can use an Authors Productivity Plot in Figure 06, showing the top 10 most productive authors over all years.

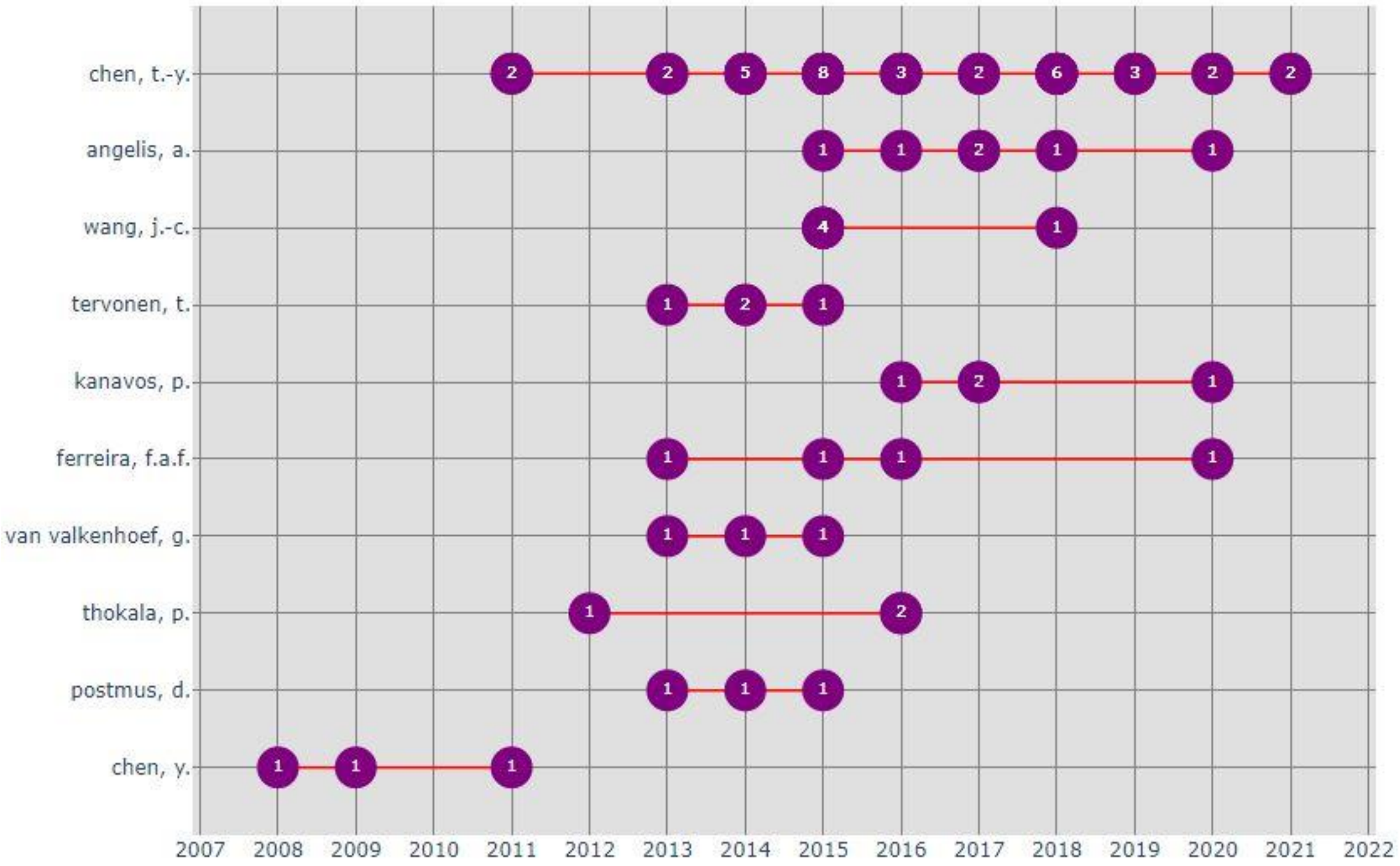

**Figure 06- Authors Productivity Plot -Top 10**

Figure 06 shows that Chen, T.-Y. has published the most documents consistently from 2011 to 2022. Also, we can highlight Angelis, A, and other authors having notable productivity. This plot is also interactive, and hovering over a node displays the document ID, providing further information about each author's publications. By examining the publications of the most prominent authors, researchers can better understand the current state of research in this field and identify areas where further investigation is needed.

Finally, we can end our EDA by verifying some extra statistics about the metadata using Bar Plots. The options that can be displayed include: documents per year, citations per year, past citations per year, Lotka's Law (LOTKA, 1926), sources per document, sources per citation, authors per document, authors per citation, authors per h-index, Bradford's Law, institutions per documents, institutions per citations, countries per documents, countries per citations, language per documents, keywords plus per documents, and authors' keywords per documents. Continuing our analysis, we can verify if document production has declined over the years, so Figure 07 shows the documents per year plot.

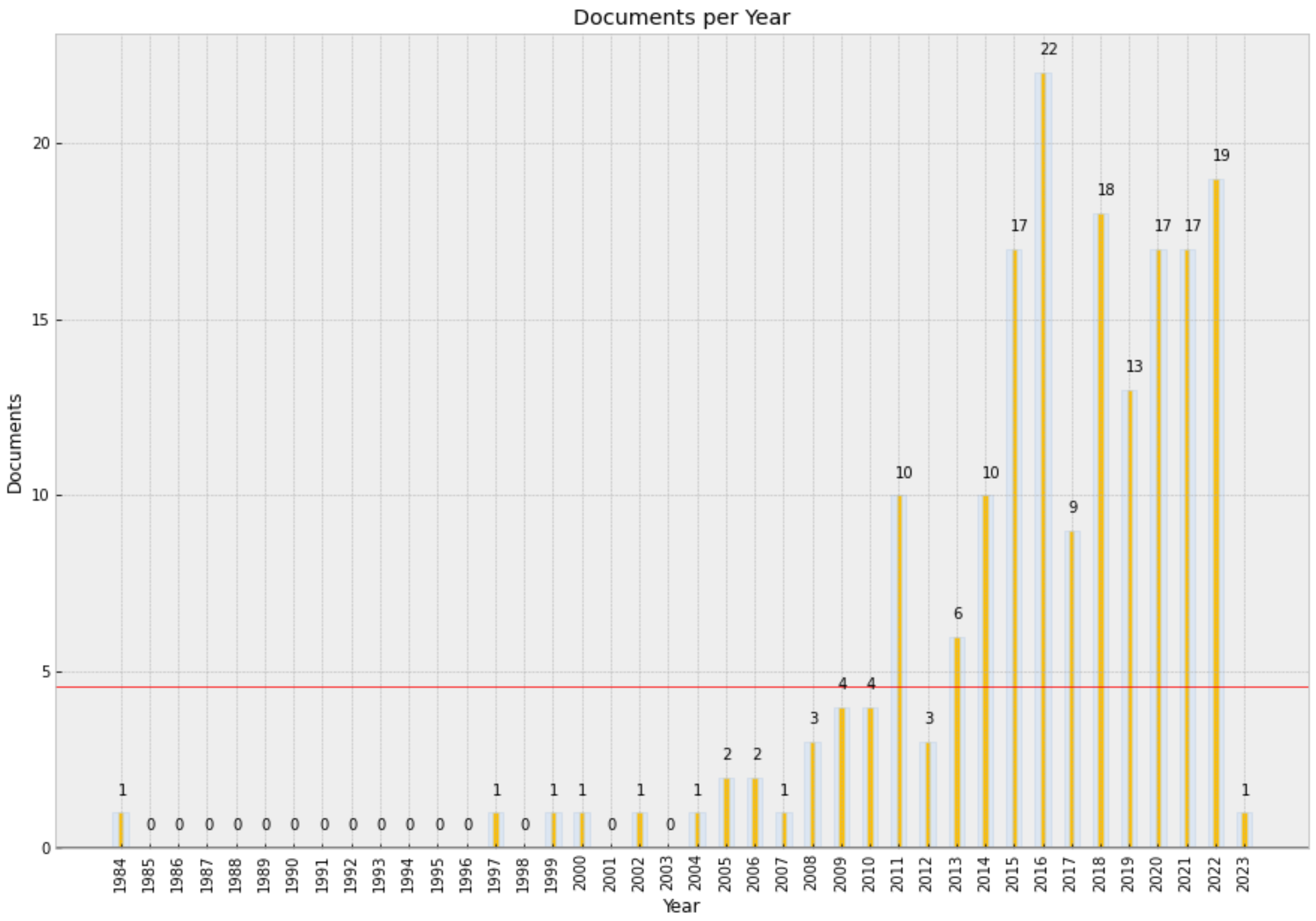

**Figure 07- Bar Plot -Documents per Year**

Figure 07 shows a positive trend in the number of documents published in the field, which indicates a growing interest in the research. It is encouraging to see that the average number of articles published per year over the entire period is almost 5, which indicates a consistent level of interest in the field. Furthermore, considering the past five years (2018-2022), the average number of documents published yearly has increased to almost 17, suggesting an even greater interest in this area. These results indicate many opportunities for researchers to contribute to the field and establish collaborations.

## 4.2 pyBibX – Network Analysis

Now that the EDA is complete, we can move on to Network Analysis, which focuses on the relationships between scientific publications, authors, and journals. In Network Analysis, publications, authors, and journals are represented as nodes, and their relationships are represented as edges. Network analysis can identify collaboration patterns between authors or institutions, map ideas within a field, or identify critical publications or journals within a particular research area (DING, 2011).

Adjacency Analysis studies relationships or connections between entities, often represented as a network or graph. In citation analysis, a type of Adjacency Analysis, the connections between documents are represented by links between nodes. Each node represents a document, and the links represent citations from one document to another. In our approach, a blue node represents each document, and red nodes represent its references. By visualizing these nodes and their connections, we can understand how the documents are related and which references are frequently cited across multiple documents. Figure 08 shows the citation analysis for a subset of documents and references cited at least four times.

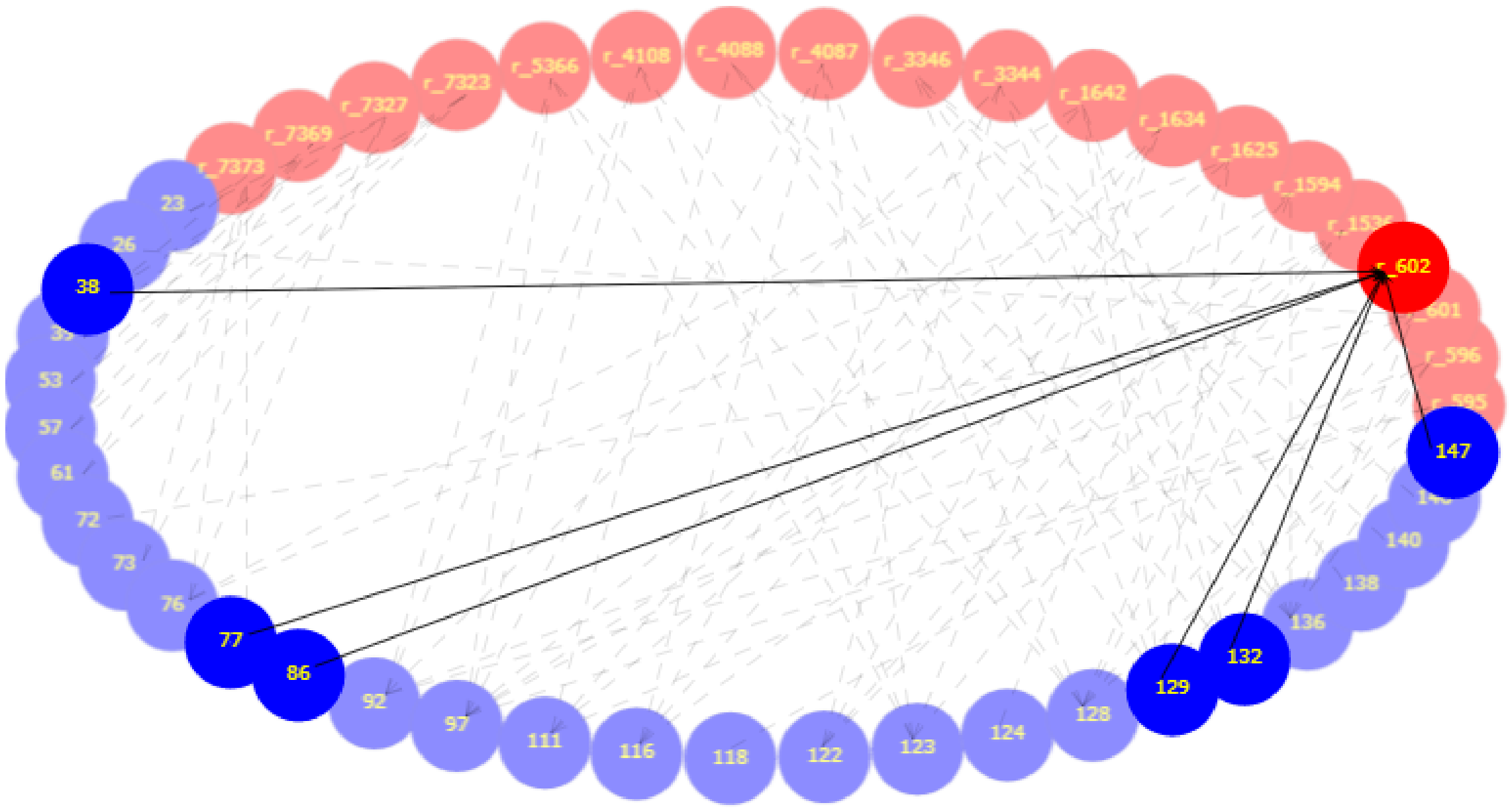

**Figure 08- Citation Analysis**

Figure 08 shows the mentioned subset of documents and references and highlights the reference "r_602" (ATANASSOV, 1986), which is cited by the documents "38" (CHEN, 2020), "77" (CHEN, 2018), "86" (ZHANG et al. 2017), "129" (CHEN, 2014a), "132" (CHEN, 2014b), and "147" (CHEN, 2011). This subset consisted of 19 references (red nodes) and 28 documents (blue nodes), which may seem relatively small compared to the total 184 documents in the dataset. However, it is critical to note that this subset was specifically chosen based on the criteria of being cited at least four times, indicating their higher significance and relevance within the field of study. The highlighted reference shows the work of Atanassov about Intuitionistic Fuzzy Sets was cited five times by Chen, T.-Y. and once by Zhang et al. Suggesting that Chen, T.-Y. may be particularly interested in this topic as he has cited the reference multiple times over the years.

The Citation History Analysis is a powerful tool for understanding a particular research field's evolution. Analyzing the citations between the 184 documents makes it possible to identify essential contributions, key authors, and emerging trends. This type of analysis can help researchers to identify gaps in the literature, potential collaborators, and new directions for their research. Moreover, it can help track a particular publication's impact over time and how it has influenced subsequent research. Figure 09 presents the Citation History Analysis, with document "97" highlighted.

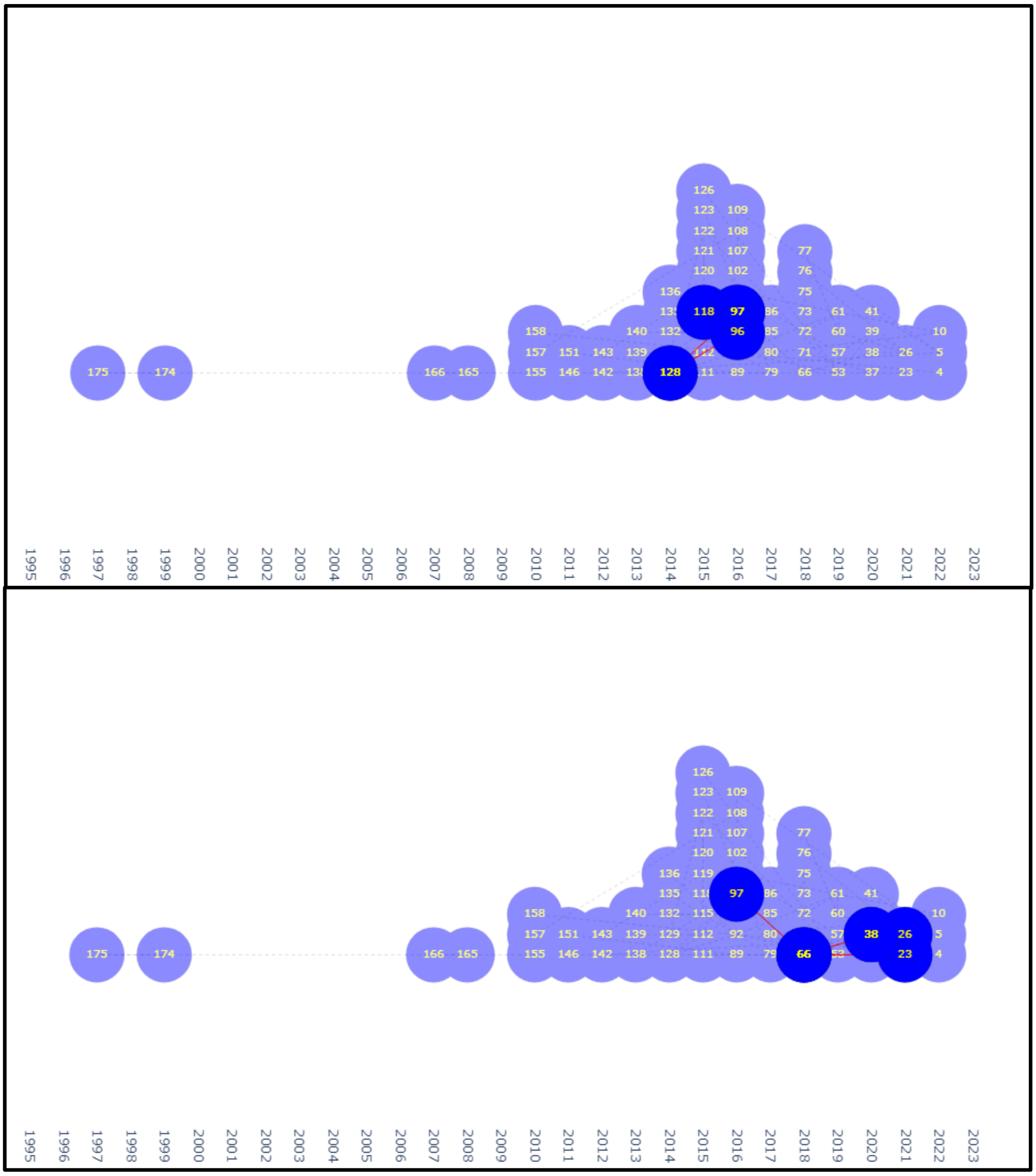

**Figure 09- Citation History Analysis**

Figure 09 depicts a fascinating network of documents cited by a central document labeled “97”. The upper graph reveals a chain reaction initiated by “97” extending to past documents. According to the network, “97” has cited “118” and “96,” while “97” and “96” have cited “128.” The lower graph showcases the same chain reaction, but this time moving towards future documents. The document labeled “66” is seen to have cited “97”. Documents “38” and “23” have cited “66”, and finally, “26” has cited “38”, thus completing the network. This comprehensive visualization of the citation network provides valuable insights into the relationships between different documents and authors, enabling researchers to draw many conclusions about the evolution of a particular field over time. By examining the citation patterns of individual works and authors, researchers can identify the collaborations and affiliations that have shaped the development of a field. In this example, all documents were works published by Chen, T.-Y.

According to our EDA, performing a Collaboration Analysis centered on the most prolific researcher, Chen, T.-Y., would be intriguing. According to Castañer and Oliveira (2020), Collaboration Analysis is a methodology used to study the interactions and relationships between individuals, groups, or organizations that collaborate toward achieving a common goal. The Collaboration Analysis results in understanding how collaboration networks are structured, how they function, and how they can be leveraged to facilitate collaboration, knowledge sharing, and innovation. Figure 10 shows the extension of the researcher Chen, T.-Y. collaboration network.

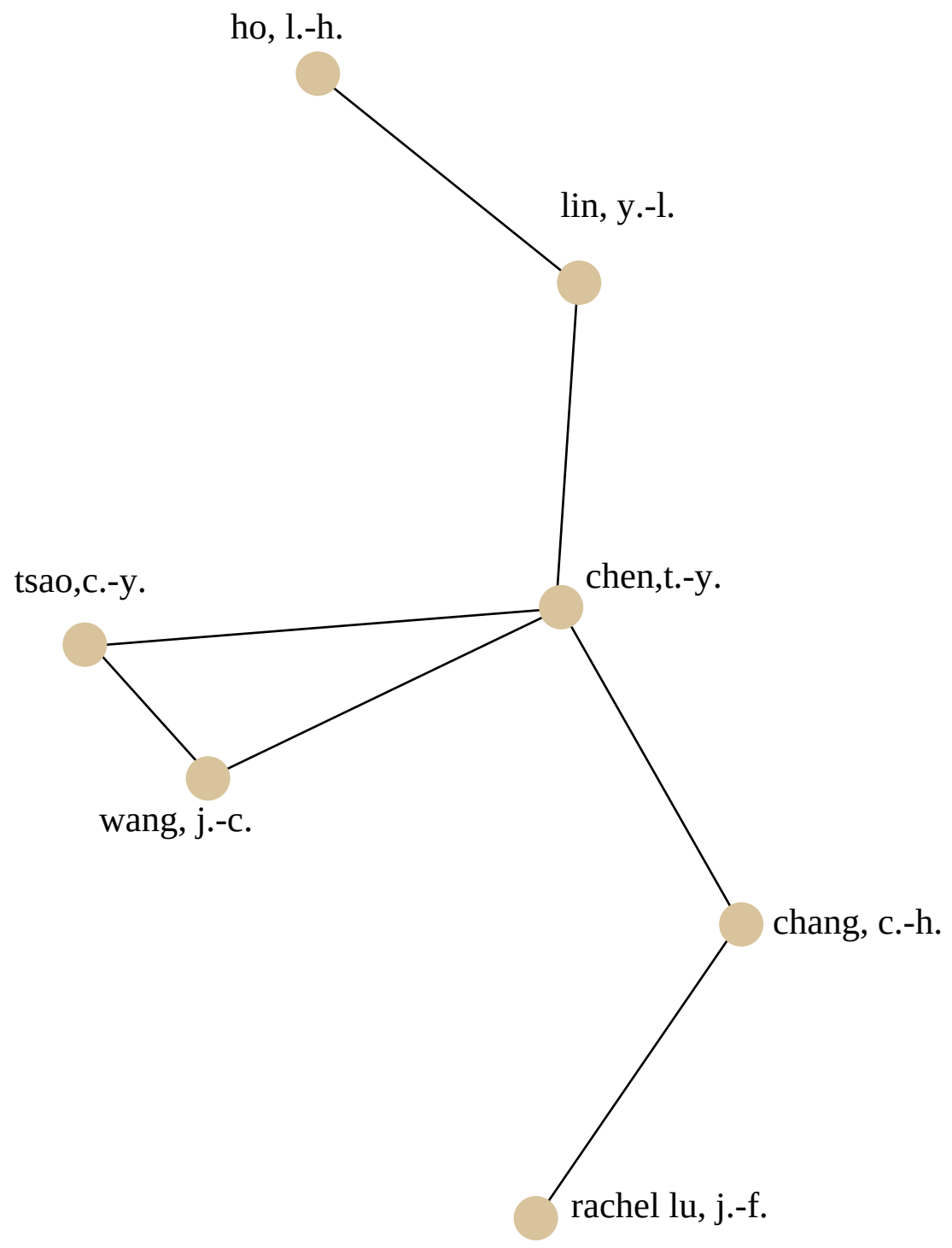

**Figure 10- Collaboration Network Analysis**

Figure 10 reveals that Chen, T.-Y. has only collaborated with four other researchers, including Lin, Y.-L., Chang, C.-H., Tsao, C.-Y., and Wang, J.-C. In addition, Tsao, C.-Y. has collaborated with Wang, J.-C., Lin, Y.-L. has collaborated with Ho, L.-H., and Chang, C.-H. has collaborated with Rachel Lu, J.-F. Because Chen, T.-Y. has written 35 documents of the 184 available, it is clear that he is the most prolific and influential researcher in this field. However, the fact that he only collaborates with four other researchers indicates that, if possible, there may be room for more collaboration and cross-disciplinary research in this field.

We can take a similar approach by shifting our focus to the documents and examining how they "collaborate" with each other. As documents cannot collaborate in the traditional sense, we need to perform a Similarity Analysis instead. Similarity Analysis is a valuable technique for exploring the relationships between different documents, and it can be conducted using various methods. One popular approach is bibliographic coupling and co-citation methods, which involve analyzing the references cited within different documents to identify patterns and relationships between them (BOYACK and KLAVANS, 2010). In our example, we will use co-citation analysis to identify documents with common references. Specifically, we will be looking for documents with at least ten common references, allowing us to identify highly related documents regarding the topics they cover and the sources they draw upon. Figure 11 represents the co-citation analysis of the 184 documents.

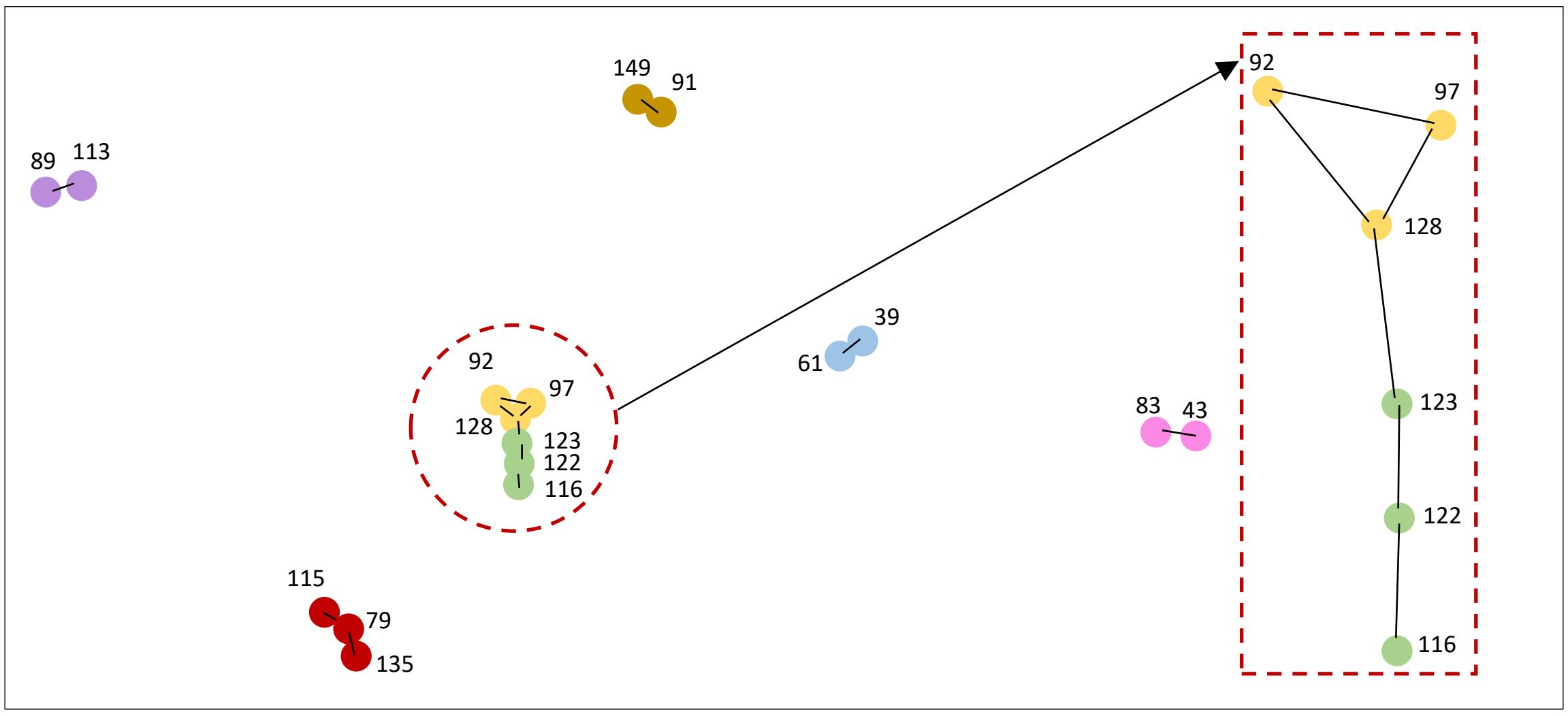

**Figure 11- Similarity Analysis – Co-citation Analysis**

In Figure 11, we can observe the emergence of seven distinct clusters labeled with colors such as purple, red, yellow, green, brown, blue, and pink. By magnifying the yellow and green clusters, we can see the link between them, which serves as a significant connection between both clusters. The purple cluster comprised documents “89” and “113”; the brown cluster included documents “149” and “91”; the red cluster had documents “115”, “79”, and “135”; the yellow cluster the documents “92”, “97”, and “128”; the green cluster the documents “123”, “122”, and “116”; the blue cluster contained documents “61” and “39”, and finally, the pink cluster had documents “83” and “43.” Interestingly, it was found that Chen, T.-Y. had authored documents in the yellow, green, blue, and red clusters, indicating this author’s broad coverage of topics.

Analyzing the yellow and green clusters, the pairs “92-97” possess 14 common references, “92-128” has 10, and “97-128” has 20. The link between the yellow and green clusters, “122-123”, has ten common references. In the green cluster, “123-122” has 26 common references, and “122-116” has 13. Based on the analysis, it can be concluded that there is a substantial similarity between the documents in the yellow and green clusters, as evidenced by the high number of shared references between them. The link between the two clusters (document pairs “122-123”) also shows moderate similarity. The green cluster, in particular, has a high level of similarity among its documents, with the pairs “123-122” and “122-116” having many common references. These findings suggest that the documents in the yellow and green clusters may share common research topics and themes.

Scientific research thrives on collaboration, and analyzing collaboration networks can offer valuable insights into how researchers and institutions work together to advance knowledge. To gain a more comprehensive understanding of how scientific knowledge is produced and shared, a World Map Collaboration Analysis can be performed to explore how researchers and institutions from different countries collaborate on research projects. Such an analysis can uncover patterns and trends in international collaborations, identify key players and institutions, and help guide funding decisions and policy-making. Figure 12 shows the World Map Collaboration Analysis.

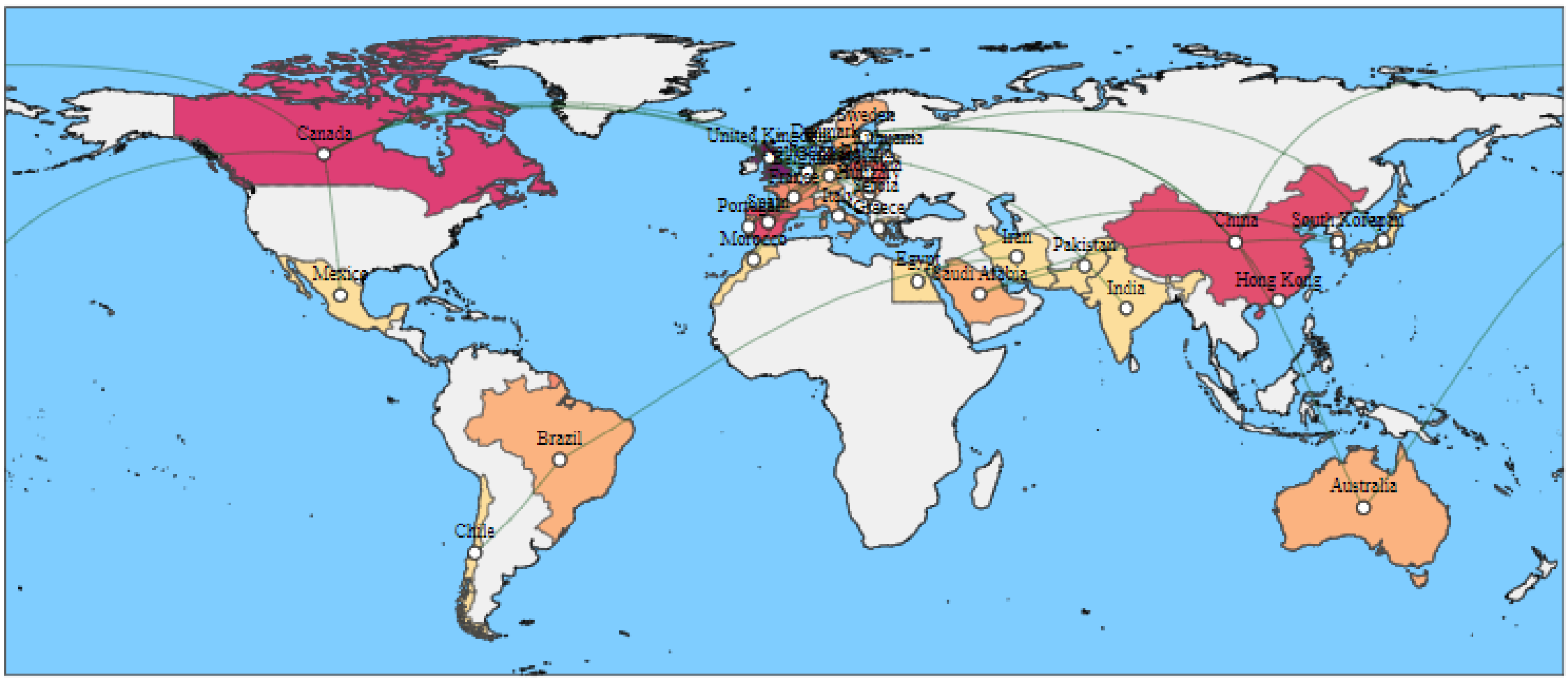

**Figure 12-World Map Collaboration Analysis**

Figure 12 shows the World Map Collaboration Analysis results, highlights the importance of collaboration in scientific research, and reveals interesting patterns in international collaboration. The graph shows that European researchers tend to collaborate more frequently with their European counterparts, indicating a solid intra-continental collaboration network. Canada and China emerged as prominent collaboration hubs, suggesting that these countries play essential roles in facilitating global scientific knowledge production. On the other hand, the analysis also revealed that countries such as Brazil, Chile, and Australia have weaker collaboration networks, which could indicate a need for increased efforts to promote collaboration in these regions. The World Map Collaboration Analysis provides valuable insights into the patterns and dynamics of international scientific collaboration, which can inform future efforts to promote collaboration and knowledge exchange across borders.

### 4.3 pyBibX – Artificial Intelligence Analysis

After performing the EDA and Network Analysis, our next step is to use Artificial Intelligence tools to gain deeper insights into the collection of documents. In particular, Embeddings, Topic Modeling, and Text Summarization, both based on Deep Learning techniques, can help to uncover hidden themes and patterns within the data. By automatically identifying and summarizing key topics and trends, these tools can provide a more efficient and accurate way to extract meaning from large amounts of text data.

Our study combined UMAP with Embeddings to project the abstracts onto a lower-dimensional space and visualize their relationship. The Embeddings are obtained using the Sentence-BERT Deep Learning Model developed by Reimers and Gurevych (2019). In addition to visualizing the relationship between abstracts, the Embeddings can be used as input to the other Deep Learning techniques. By using these Embeddings as input, the Deep Learning models can better capture the semantic meaning and context of the text. These representations can improve results in tasks such as identifying key topics or generating concise summaries, demonstrating the versatility and usefulness of Embeddings in various NLP tasks.

Topic Modeling is a method that automatically detects the main themes and patterns hidden in a set of documents. Analyzing word and phrase patterns within texts identifies groups of words that frequently co-occur, which are then interpreted as topics, representing the underlying concepts or ideas in the collection of documents. The primary purpose of Topic Modeling is to offer a means of identifying and comprehending the most significant themes within a vast amount of text data. The results of the Topic Modeling using the BerTopic Deep Learning Model developed by Grootendorst (2022) are presented in Table 05.

**Table 05 – Topic Modeling**

| | Associated Words | Topic Label | Central Document |
|---|---|---|---|
| Topic 0 (153) | type, selection, ranking, data, sets, preference, process, information, interval, fuzzy | Decision-making and Preference Analysis | “96” (CHEN, 2016) |
| Topic 1 (31) | benefit, impact, treatment, assessment, patients, risk, framework, care, hta, health | Health Technology Assessment and Patient Care | “74” (ANGELIS, 2018) |

Based on the results presented in Table 05, it can be concluded that Topic Modeling has successfully identified the two main topics in the 184 documents. Topic 0 is the most prominent, with a majority of 153 documents, while Topic 1 has only 31 documents. The words associated with Topic 0 suggest a focus on “Decision-making and Preference Analysis”, a topic label supported by the set of words that helps characterize this topic. The central document, “96” by Chen (2016), mainly represents the topic theme. On the other hand, Topic 1 seems to focus on “Health Technology Assessment and Patient Care”, as indicated by its associated words. The central document is “74” by Angelis (2018). These findings suggest that the collection of documents is dominated by decision-making and preference analysis studies, with a smaller subset of studies related to health technology assessment and patient care. Additionally, these findings are supported by the evidence presented in Figure 02.

The central documents identified, “96” by Chen (2016) and “74” by Angelis (2018), can be summarized using Abstractive and Extractive Summarization techniques. Abstractive Summarization involves creating a new summary that conveys the essential information from the original text. In contrast, extractive Summarization involves selecting the most significant sentences or phrases from the original text to create a condensed version. The primary difference between these two techniques is that Abstractive Summarization requires the creation of new text that accurately conveys the meaning of the original. On the other hand, Extractive Summarization involves selecting and condensing text from the original. Abstractive Summarization can be more challenging but may capture the whole meaning of the original text, while Extractive Summarization is more straightforward but may not convey the whole meaning of the original text (YADAV et al., 2022).

Both summarization models utilized advanced techniques; precisely, the Extractive Summarization was accomplished using a BERT model crafted by Derek (2019). While BERT has indeed achieved state-of-the-art results on numerous NLP tasks, it is essential to recognize that fine-tuning may be necessary for optimal performance in documents with particular domains. Fine-tuning BERT on domain-specific corpora can better capture the nuances, terminology, and context of the specialized subject matter. This tailored approach ensures BERT’s more accurate and practical application to domain-specific tasks. This results in improved performance and a greater understanding of the unique characteristics inherent to the particular domain.

On the other hand, Abstractive Summarization was accomplished by utilizing PEGASUS, a sophisticated method developed by Zhang et al.(2019), and chatGPT (OPENAI, 2023), the current state-of-the-art for Large Language Models (LLM), and not always freely accessible. The outcomes of the techniques are presented in Table 06, with both abstracts being utilized as input. The original abstracts of the articles can be conveniently accessed on their respective scientific journals’ websites by referencing their unique DOI numbers.

**Table 06 – Text Summarization**

| Central Document | Abstractive Summarization - PEGASUS |
|---|---|
| "96" (CHEN, 2016) | The author is affiliated with Vilnius Gediminas Technical University (VGTU) [sic] |
| "74" (ANGELIS, 2018) | The use of multiple criteria decision analysis in health technology assessment has been identified as a prospective methodology for assisting decision makers in evaluating the benefits of new medicines in health technology assessment [sic] |
| **Central Document** | **Abstractive Summarization - chatGPT** |
| "96" (CHEN, 2016) | Interval-valued intuitionistic fuzzy sets (IVIFSs) provide a way of addressing uncertain and ambiguous properties. IVIFSs are compared using an inclusion comparison possibility, which results in an inclusion-based index. This index is used to construct a linear programming model (for consistent information) and an integrated, nonlinear programming model (for inconsistent information) to estimate the criterion weights and the optimal ranking order of the alternatives under an incomplete preference structure. The proposed method is illustrated by a [sic] |
| "74" (ANGELIS, 2018) | Multiple criteria decision analysis (MCDA) is a methodology that has been identified as a prospective way to assist decision makers in evaluating the benefits of new medicines in health technology assessment (HTA). However, there is limited empirical evidence from real-world applications to test this methodology. In order to test the MCDA methodology in a real-world setting, four experts participated in a decision conference where they used the Advance Value Framework to evaluate the benefits of three different treatments for metastatic castrate-resistant [sic] |
| **Central Document** | **Extractive Summarization - BERT** |
| "96" (CHEN, 2016) | Abstract: The theory of interval-valued intuitionistic fuzzy sets provides an intuitive and feasible way of addressing uncertain and ambiguous properties. Many useful models and methods have been developed for multiple criteria decision analysis within the interval-valued intuitionistic fuzzy environment. [sic] |
| "74" (ANGELIS, 2018) | Background. Multiple criteria decision analysis (MCDA) has been identified as a prospective methodology for assisting decision makers in evaluating the benefits of new medicines in health technology assessment (HTA); however, limited empirical evidence exists from real-world applications. Measurements. Participants' value preferences were elicited involving criteria selection, options scoring, criteria weighting, and their aggregation. Conclusion. [sic] |

Table 06 presents the Abstractive and Extractive summaries of the central documents "96" by Chen (2016) and "74" by Angelis (2018). In comparing the original abstracts to the summaries, it is clear that PEGASUS and BERT summarizations have missed important information presented in the original abstracts. chatGPT's Abstractive summaries provide a more coherent and concise representation of the original abstracts.

Regarding Chen's work, the Abstractive Summarization from PEGASUS only provided information about the author's affiliation, which is not relevant to the content of the document. On the other hand, chatGPT Abstractive summary captures the essence of the original abstract, discussing the use of interval-valued intuitionistic fuzzy sets, inclusion comparison possibility, and the development of linear and nonlinear programming models to estimate criterion weights and optimal ranking order. The Extractive Summarization highlighted the Interval-valued Intuitionistic Fuzzy Sets theory and its usefulness in MCDA. However, it doesn't provide a comprehensive overview of the proposed models' methodology or application.

For Angelis's work, the Abstractive Summarization from PEGASUS only provides a general statement about using MCDA in HTA. Chatgpt Abstractive Summarization is informative and covers essential aspects of the study, such as MCDA's potential in HTA, the real-world testing of the Advance Value Framework, and the evaluation of three different treatments for metastatic castrate-resistant prostate cancer. At the same time, the Extractive Summarization touches on the study's background, measurements, and conclusion, highlighting MCDA as a prospective methodology in HTA. However, it lacks details about the Advance Value Framework and the specific treatments being evaluated.

Although PEGASUS and BERT summarization techniques have not yet achieved the same level of proficiency as chatGPT, it is essential to underscore the notable difference when utilizing chatGPT within applications, as it requires a paid API key, or a subscription to access more advanced models. At the same time, PEGASUS and BERT remain freely accessible, and this fact renders PEGASUS and BERT more inclusive for a broader spectrum of users and researchers. Furthermore, with the ongoing evolution of these techniques, it is conceivable that their performance may eventually rival or even surpass chatGPT. As Artificial Intelligence techniques rapidly progress, these freely available summarization methods hold immense potential for development and refinement, eventually becoming invaluable resources for those unable or unwilling to invest in paid alternatives like chatGPT.

In this very last step, we present the seamless integration of chatGPT with our library, made possible by custom-designed functions that streamline the examination and interpretation of various bibliometric results. These include EDA Reports, WordClouds, N-Grams, Evolution Plots, Sankey Diagrams, Author Productivity Plots, Bar Plots, Citation Analysis, Collaboration Analysis, Similarity Analysis, and World Map Collaboration Analysis. The fusion of human inquisitiveness and chatGPT has the potential to enhance comprehension and reveal novel perspectives in traditional bibliometric methodologies.

Taking Figure 07 (Bar Plot of documents produced by year) as an illustration, after generating the bar plot, the researcher can pose any questions that can aid in better interpreting the results. We will demonstrate this approach by sequentially conducting a series of distinct queries.

Query 1: *Give me insights about the given information*

Response: *The data shows that there has been an overall trend in increasing documents from 1984 to 2022. Between 1984 and 1995, there was not much fluctuation as the value for documents was almost 0 for all years except 1984 and 1997. Between 1996 and 2021, there was a slow but steady increase in documents that was more evident in the last decade (2010-2020). The highest number of documents was recorded in 2016 and the second highest in 2020. The documents dropped in 2017 to 9 and increased again in 2018 and 2019 to 18 and 13 respectively. In 2021, the documents increased again to 17. It looks like documents are expected to remain at this level in the upcoming years*. [sic]

Query 2: *What happened before and after the year 2004?*

Response: *Before 2004, the number of documents released per year was quite low, with most years seeing no new documents. The year 2004 saw the release of one new document. After 2004, the number of documents per year increased steadily, with 2005 and 2006 each seeing two documents released and 2008 seeing three. The number of documents released continued to increase steadily, with a peak of 22 documents released in 2016. Since then, the number of documents released has declined slightly, but still remained relatively high compared to before 2004.* [sic]

Query 3: *Give me a list of ways to improve the research interest in this topic.*

Response: *1. Analyze the trends in the data to determine the factors that are driving or hindering the growth of research interest. 2. Identify potential collaborators who can help expand the scope of the research. 3. Utilize social media and other digital platforms to promote the research and attract more attention. 4. Develop a comprehensive and engaging research proposal to outline the project's goals and objectives. 5. Create an outreach strategy to engage with the general public and promote awareness about the research. 6. Leverage existing research and data to inform the research and provide additional contexts. 7. Develop a comprehensive data collection and analysis plan to capture the necessary information. 8. Present the data in visually appealing and comprehensive formats to better communicate the findings. 9. Publish research papers in prestigious journals to increase the visibility of the research. 10. Participate in relevant events and conferences to interact with other researchers and discuss ideas.* [sic]

The three queries posed to chatGPT focused on extracting insights from the data regarding the documents produced per year. The initial query sought a broader understanding of the information. In contrast, the latter query specifically requested details about the years preceding and succeeding 2004. Finally, the third query inquired about a compilation of methods to bolster interest in the research topic under examination. The responses generated by chatGPT were of good quality, offering lucid and succinct insights into the data and providing valuable suggestions for enhancing research interest. Nevertheless, there is potential for enhancing the responses, particularly between the first two queries, as the second response predominantly reiterates the content from the first response. One way to achieve that could be by crafting more intricate queries encouraging chatGPT to delve deeper into the data and explore novel aspects, generating more diverse and informative responses.

Its worth noting that we have simulated how chatGPT has a remarkable capacity to ignite the curiosity of humans, as evinced by the progressive nature of the questions presented in the three queries. From the first query, there is a pursuit for a generalized understanding of the data, and this provoked an explicit interest in a specific year, leading to the second query. And as the curiosity flourished, the third query represented the solicitation for prospective guidance in the subject matter. Throughout this cyclical process that could be continued with more profound ramifications, chatGPT skillfully navigated the researcher's inquiries, showcasing its ability to cultivate intellectual curiosity and enable a deeper exploration of the underlying data.

### 4.4 pyBibX – Discussion

The theoretical applications of pyBibX, in light of its research objectives, span multiple dimensions that contribute to the advancement of bibliometric analysis understanding and its vital role in sculpting the scientific landscape. By incorporating advanced Artificial Intelligence techniques, pyBibX refines the methodological sophistication of bibliometric analysis, paving the way for innovative approaches. In other words, pyBibX enhances the theoretical foundations of bibliometrics by using Artificial Intelligence techniques and Artificial Intelligence assisted conversational tools in enabling a new type of synergy never seen before. Simultaneously, the emphasis on visual elegance deepens data comprehension and fosters the development of visualization theories and best practices for effective communication in scientific research.

It is worth noting that pyBibX enhances the results of EDA and Network Analysis by integrating advanced Artificial Intelligence capabilities into its code, such as Topic Modeling, Embedding Vectors, Text Summarization, and general NLP tasks. This fusion of Artificial Intelligence capabilities boosts the accuracy and efficiency of bibliometric analysis, uncovering unique patterns and trends that traditional methods might miss. In practical terms, pyBibX streamlines the bibliometric analysis process by reducing the manual effort and time expenditure typically associated with such tasks. This increased efficiency enables scientists to respond more swiftly to new findings and developments in their respective fields.

A distinguishing feature of pyBibX is its ability to facilitate engagement between researchers and chatGPT, enabling them to pose questions and receive comprehensive answers based on the analysis findings. This interactive aspect deepens the understanding of the data and hastens the discovery of concealed knowledge, ultimately promoting scientific research progress. With its innovative features, pyBibX provides researchers with a robust and insightful tool for bibliometric analysis.

As the scientific community expands and amasses immense volumes of data, the demand for proficient and refined bibliometric analysis tools rises in importance. PyBibX responds to this necessity by offering a state-of-the-art, Artificial Intelligence-driven solution enabling researchers to explore complex datasets deftly. Its pioneering amalgamation of advanced Artificial Intelligence capabilities furnishes researchers with the means to remain at the vanguard of their disciplines, unveiling concealed connections and contributing to the collective knowledge base.

## 5. Conclusion

The pyBibX library stands as a veritable asset for the academic sphere, offering invaluable advantages in analyzing scientific data, discerning patterns and trends, and generating insightful visualizations that foster collaboration. The library's efficacy resides in its capacity to extract meaningful insights from complex data, automate specific tasks, and maintain accuracy and consistency in data analysis, thereby mitigating the risk of human error. Our bibliometric investigation encompassed a multifaceted approach, integrating EDA, Network Analysis, and Artificial Intelligence Analysis through employing Deep Learning methodologies such as Embedding, Topic Modeling, and Text Summarization with cutting-edge tools like BERT, Sentence-BERT, BERTopic, Pegasus, and chatGPT. The outcomes of each technique yield substantial time and resource savings for researchers, enabling them to concentrate on elevated tasks such as hypothesis formulation, experimental design, and manuscript composition. In a nutshell, pyBibX embodies an instrument for the scholarly community that can enrich the analysis and synthesis of bibliographic information.